\documentclass[useAMS,usenatbib]{mn2e}
\usepackage{graphicx,amssymb,color}
\usepackage[normalem]{ulem}

\newcommand{\forref}{ }
\newcommand{\rev}{ }

\title[GB Yarkovsky asteroid-planet drift]
{Speeding past planets? Asteroids radiatively propelled by giant branch Yarkovsky effects}

\author[Veras, Higuchi \& Ida]{
Dimitri Veras$^{1,2}$\thanks{E-mail: d.veras@warwick.ac.uk}\thanks{STFC Ernest Rutherford Fellow},
Arika Higuchi$^{3}$,
Shigeru Ida$^{4}$
\\
$^{1}$Centre for Exoplanets and Habitability, University of Warwick, Coventry CV4 7AL, UK
\\
$^{2}$Department of Physics, University of Warwick, Coventry CV4 7AL, UK
\\
$^{3}$RISE Project Office, National Astronomical Observatory of Japan,
Osawa, Mitaka, Tokyo 181-8588, Japan
\\
$^{4}$Earth-Life Science Institute, Tokyo Institute of Technology, Meguro, Tokyo 152-8550, Japan
}

\pubyear{2019}

\begin{document}
\label{firstpage}
\pagerange{\pageref{firstpage}--\pageref{lastpage}}
\maketitle

\begin{abstract}
Understanding the fate of planetary systems through white dwarfs which accrete
debris crucially relies on tracing the orbital and physical
properties of exo-asteroids during the giant branch phase of stellar evolution.
Giant branch luminosities exceed the Sun's by over three orders of magnitude, leading
to significantly enhanced Yarkovsky and YORP effects on minor planets. Here, we place 
bounds on Yarkovsky-induced differential migration between asteroids and planets 
during giant branch mass loss by modelling one exo-Neptune
with inner and outer exo-Kuiper belts. In our bounding models, the asteroids move too
quickly past the planet to be diverted from their eventual fate, which can range from: 
(i) populating the outer regions
of systems out to $10^4-10^5$~au, (ii) being engulfed within the host star, or (iii) experiencing
Yarkovsky-induced orbital inclination flipping without any Yarkovsky-induced semimajor
axis drift.  {\rev In these violent limiting cases, temporary resonant trapping of asteroids with 
radii of under about $10$ km by the planet is insignificant, and capture within the planet's 
Hill sphere requires fine-tuned dissipation}.
The wide variety of outcomes presented here
demonstrates the need to employ sophisticated structure and radiative exo-asteroid models in future
studies. Determining where metal-polluting asteroids reside around a white
dwarf depends on understanding extreme Yarkovsky physics.
\end{abstract}


\section{Introduction}

Upon leaving the main sequence and ascending the giant branch, a star which hosts a 
planetary system will subject its
orbiting constituents to a violent cocktail of destructive forces. These include (i) physical
expansion of the star's envelope to au-scale distances, (ii) shedding of mass through superwinds,
and (iii) luminosities which reach between $10^3L_{\odot}$ and $10^4L_{\odot}$. The dynamical and
physical consequences for giant planets, terrestrial planets, minor planets, boulders, 
pebbles and dust are only starting to be characterised \citep{veras2016a} at a sufficiently-detailed
level to connect with the abundant data available in white dwarf planetary systems.

\subsection{Giant branch effects}

An expanding giant branch star can engulf objects which orbit too closely. Determining
the critical engulfment distance has already been the subject of intense scrutiny 
\citep{villiv2009,kunetal2011,musvil2012,adablo2013,norspi2013,lietal2014,
viletal2014,madetal2016,staetal2016,galetal2017}, and is strongly dependent on both the tidal
prescription and stellar model that one adopts. In general, however, this critical engulfment
distance resides at a few au, and is greater for gas giants than for terrestrial planets.
In the solar system, Mercury and Venus will easily be engulfed, Mars will not, and the outcome
for the Earth remains unclear \citep{schcon2008}.

Giant branch stellar mass loss will precipitate orbital changes for all types of planets. In general, even 
for isotropic mass loss, a planet's eccentricity, semimajor axis and argument of pericentre
will vary \citep{omarov1962,hadjidemetriou1963}\footnote{Anisotropic mass loss generates changes in all
orbital elements \citep{veretal2013a,doskal2016a,doskal2016b}.}. However, for planets and asteroids
within a few hundred au of their parent star, the adiabatic approximation may be employed \citep{veretal2011},
where orbital eccentricity is effectively conserved and semimajor axis expansion scales simply with stellar mass loss.
Nevertheless, despite the invariance of the semimajor axis ratios of multiple objects in the adiabatic approximation, the central mass alterations can incite instability 
\citep{debsig2002,bonetal2011,debetal2012,veretal2013b,voyetal2013,frehan2014,musetal2014,vergae2015,
veretal2016,veras2016b,musetal2018,smaetal2018}, which is manifest both along the giant branch and white 
dwarf phases. In the solar system, the giant planets are separated sufficiently far from one another
to prevent post-main-sequence instability, unless there exists an additional Planet Nine-like companion 
\citep{veras2016b}.

The time-varying giant branch luminosities affect planets and asteroids differently. Planetary atmospheres are subjected
to partial or complete evaporation \citep{livsok1984,neltau1998,soker1998,villiv2007,wicetal2010}, and
their geology and surface processes are likely exposed to (as-yet-unmodelled) transformative events.
Asteroids, however, are small enough to be exposed to interior water depletion \citep{malper2016,malper2017a,malper2017b} 
and being radiatively ``pushed'' by the Yarkovsky effect
\citep{veretal2015} and spun-up through the YORP effect \citep{veretal2014a}.
These latter two phenomena have been
observed in the solar system \citep[e.g. see Fig. 1 of][]{poletal2017} but are enhanced during
the giant branch phases. In fact, the YORP effect could easily destroy 100m - 10km asteroids 
within about 7 au of their parent star through rotational fission. Giant branch luminosity also 
creates a drag force for pebbles and boulders \citep{donetal2010}, which becomes important
for objects which are too small for the Yarkovsky effect to play a role in determining the final 
orbital state \citep{veretal2015}.

\subsection{White dwarf planetary systems}

After the star has become a white dwarf, the mutual perturbations amongst the remaining planets, 
asteroids, pebbles and fragments conspire to thrust enough material into the white dwarf photosphere
to be observable in one-quarter to one-half of all observed white dwarfs 
\citep{zucetal2003,zucetal2010,koeetal2014}. One of these exo-asteroids is currently in the process of
disintegrating, a phenomenon which has been seen in real time on a nightly basis for the last several years \citep{vanetal2015}. Such 
disintegration produces debris
discs \citep{jura2003,jura2008,debetal2012,veretal2014b}, of which about 40 are now known \citep{farihi2016}.
Another exo-asteroid, one with high internal strength, has been found embedded inside one of these
discs \citep{manetal2018}. The importance of these minor bodies is highlighted by how their
innards after break-up are regularly observed in white dwarf photospheres
\citep{kleetal2010,kleetal2011,gaeetal2012,juretal2012,xuetal2013,xuetal2014,juryou2014,wiletal2015,
wiletal2016,xuetal2017,haretal2018,holetal2018}. These observations provide the most direct and 
extensive known measurements
of the bulk chemical composition of the building blocks of exoplanets.
 
\subsection{Plan for paper}

Here, we seek to achieve a better understanding of the post-main-sequence orbital distribution of exo-asteroids
by (i) considering their Yarkovsky-induced interactions with a planet during the giant branch phase of
evolution, and (ii) applying numerical Yarkovsky models which can place bounds on possible behaviours. 
Because Yarkovsky acceleration affects only exo-asteroids and not exo-planets, the
resulting relative orbital evolution at first glance might appear similar to convergent or divergent
migration between two bodies within a protoplanetary disc. In Section 2, we detail the limitations of
this analogy, which provides context for our modelling efforts.
Sections 3 and 4 then, respectively, establish our numerical simulations and reports on their outcomes. 
We discuss our findings in Section 5, and summarize in Section 6.

\section{Analogy with protoplanetary disc migration}

In this section, we explore the possibility that the giant branch-induced Yarkovsky
drift can be modelled, or, at least contained, by the formalisms which have been developed
for migration within protoplanetary discs.

\subsection{Capture into resonance}

The architectures of planetary systems on the main-sequence arise from the movement of and interactions
between protoplanets within their birth disc. The prevalence of multiple planets which are currently observed
to reside close to a mean motion commensurability suggest that they drifted within the disc into such configurations.
This idea has been borne out by dozens of theoretical studies, which have characterised the relative
drift as ``convergent migration'' or ``divergent migration''.

\subsubsection{Previous modelling efforts}

In order to investigate this type of migration, several authors have solved a set of differential equations
involving $N$-body point-mass interactions, plus some perturbative accelerations (e.g. equations 17-19 of
\citealt*{terpap2007}, equation 7 of \citealt*{ogiida2012}, equation 1 of \citealt*{ogikob2013}, 
and equations 54-58 of \citealt*{teyter2014}). These equations do not
actually incorporate the potential of the disc, but rather damping timescales which mimic dissipative
disc effects. These timescales range from basic linear approximations \citep{leepea2002}
to empirical fits based on recent hydrodynamic simulations \citep{crenel2008,xuetal2018}.

\subsubsection{Types of resonances}

Capture models have focused almost exclusively on eccentricity-based
first and second-order mean motion resonances, assuming that at least one planet is on a 
near-circular orbit\footnote{Notable exceptions
  are \cite{luan2014}, who quantified
  the probability that the $4$:$2$ inclination-based mean motion resonance between Mimas and Tethys
  was the consequence of convergent migration, and \cite{elmetal2017}, who considered
  capture of massless particles in the first-order corotation eccentric resonance.}.
Assume that a first-order resonance is described as a $p+1$:$p$ mean motion resonance,
where $p$ is a positive integer. As the value of $p$ increases, and 
the configuration approaches a co-orbital state, 
the probability of capture decreases. Nevertheless, \cite{dosetal2015}
illustrated how planets can bypass -- or migrate through -- the $2$:$1$ and
$3$:$2$ mean motion resonances to become
trapped in the $4$:$3$ mean motion resonance. Similarly, figures 9 and 7 of 
\cite{papszu2005} respectively illustrated how entrapment
into the $6$:$5$ and $9$:$8$ mean motion resonances may be possible; 
\cite{quietal2013} warned, however, that capture into a 
mean motion resonance such as the $7$:$6$ requires fine tuning of parameters. 
Known moons represent good examples of bodies which may be
trapped in high-$p$ mean motion resonances: Desdemona and Portia are near 
the $13$:$12$ mean motion resonance and Cressida and 
Desdemona are near the $47$:$46$ mean motion resonance \citep{quifre2014}.

\subsubsection{Conditions for capture}

Only if the migration is convergent, slow enough, minimally stochastic, and relies on near-circular orbits 
can the planets become trapped in mean motion resonances. Each of these criteria has 
already been significantly
vetted in the literature, and were bolstered by recent revisitations of mean motion resonance capture theory
\citep{quillen2006,muswya2011,golsch2014,batygin2015,decbat2015}. Turbulence, Brownian motion and
stochasticity in general
may prevent resonant capture from occurring \citep{murchi2006,adalau2008,reipap2009,reipap2010}. Orbital 
eccentricities which exceed a few hundredths also can prevent capture, a result which has been 
demonstrated by hydrodynamical simulations \citep{hanale2018}. A final barrier to capture is migration 
speed: generally, if a planetary body crosses
the libration width of a resonance faster than the libration timescale, then capture can
be avoided \citep[e.g.][]{pansch2017}. However, capture is probabilistic, and largely depends on 
angular variables and their time evolution \citep[e.g.][]{foletal2014}.

The appendix of \cite{petetal2013} provides particularly useful equations in physical units 
for the critical migration speed and eccentricity needed to avoid certain capture for 
first-order $p+1$:$p$ resonances. They assume one body
is a planet and another is a test particle, and use $p$ in a different
way than us, allowing the integer to take on positive or negative values for resonances 
where their test particle (e.g. asteroid) is respectively interior or exterior to the planet.  
We instead use positive values for both interior and
exterior resonances, such that at a resonance the ratio $\alpha$ of the inner to outer 
semimajor axes is

\begin{equation}
\alpha \equiv \frac{a_{\rm inner}}{a_{\rm outer}} = \left(\frac{p}{p+1}\right)^{2/3}    < 1
.
\label{alphaeq}
\end{equation}

\cite{petetal2013} illustrated in their appendix that the critical speed beyond which capture may be avoided in 
terms of the planet's mean motion $n_{\rm pl}$ relative to the asteroid, or vice versa, is

\begin{equation}
\frac{dn_{\rm crit}}{dt} \approx \frac{5 \cdot 3^{5/3}}{8} \left|p^2 \left(p + 1\right) \right|^{\frac{1}{9}}n_{\rm pl}^2 \left| \frac{M_{\rm pl}}{M_{\star}} f_p \right|^{\frac{4}{3}}
.
\label{critn}
\end{equation}

\noindent{}Correspondingly, if the asteroid is external to the planet, then its critical decay rate is

\begin{equation}
\frac{da_{\rm crit}^{\rm exterior}}{dt} \approx -\frac{5 \cdot 3^{2/3}}{4} 
                                                    \sqrt{\frac{G M_{\star}}{a_{\rm pl}}}
                                                   \left(p+1\right)^{16/9} p^{-13/9}
              \left| \frac{M_{\rm pl}}{M_{\star}} f_p \right|^{\frac{4}{3}}
.
\label{dadtcritext}
\end{equation}

\noindent{}Instead, if the asteroid is internal to the planet, then its critical growth rate is

\begin{equation}
\frac{da_{\rm crit}^{\rm interior}}{dt} \approx \frac{5 \cdot 3^{2/3}}{4} 
                                                    \sqrt{\frac{G M_{\star}}{a_{\rm pl}}}
                                                   \left(p+1\right)^{-14/9} p^{17/9}
              \left| \frac{M_{\rm pl}}{M_{\star}} f_p \right|^{\frac{4}{3}}
.
\label{dadtcritint}
\end{equation}

\noindent{}For either type of resonance, the critical asteroid eccentricity beyond which
capture may be avoided is

\begin{equation}
e_{\rm crit} = \frac{2^{1/2} 3^{1/6}}{\left|p^2 \left(p + 1\right) \right|^{\frac{2}{9}}}
             \left| \frac{M_{\rm pl}}{M_{\star}} f_p \right|^{\frac{1}{3}},
\label{crite}
\end{equation}

\noindent{}where we have assumed a numerical coefficient in equation (\ref{critn}) from \cite{friedland2001}.
In the equations, $M_{\rm pl}$ is the mass of the planet, $M_{\star}$ is the mass of the star, $a_{\rm pl}$ is the 
semimajor axis of the planet, and the quantity $f_p$ is a 
function extracted from Appendix B of \cite{murder1999} that depends on whether the asteroid is interior or exterior to the planet.

For first-order resonances, in the interior case, where there is an asteroid internal to the planet, 

\[
f_{p}^{\rm interior} \equiv f_{27} + f_{\rm e}
\]

\noindent{}where $f_{\rm e} = 0$ and

\begin{equation}
  f_{27}= -\left(p + 1 + \frac{1}{2} \frac{d}{d \log{\alpha}} \right) b_{1/2}^{\left(p+1\right)}
\end{equation}

\noindent{}such that

\begin{equation}
f_{p}^{\rm interior} =-\left(p+1\right) b_{1/2}^{\left(p+1\right)} - \frac{\alpha}{4} 
\left( b_{3/2}^{\left(p\right)} - 2 \alpha b_{3/2}^{\left(p+1\right)} + b_{3/2}^{\left(p+2\right)}     \right)
.
\end{equation}

\noindent{}In the exterior case, where there is an asteroid external to the planet,

\begin{equation}
f_{p}^{\rm exterior} \equiv \alpha f_{31} + f_{\rm i}
\end{equation}

\noindent{}where $f_{\rm i} = -  \delta_{p,1}/2$ and

\begin{equation}
\alpha f_{31} =  \alpha\left(p + \frac{3}{2} + \frac{1}{2} \frac{d}{d \log{\alpha}} \right) 
                    b_{1/2}^{\left(p\right)} 
\end{equation}

\noindent{}such that

\[
f_{p}^{\rm exterior}  = \alpha\left(p + \frac{3}{2} \right) 
                    b_{1/2}^{\left(p\right)} - \frac{\delta_{p,1}}{2} 
\]

\begin{equation}
\ \ \ \ \ \ \ \ \ \ \ \ \  +\frac{\alpha^2}{4} \left(  b_{3/2}^{\left(p-1\right)} - 2 \alpha b_{3/2}^{\left(p\right)} + b_{3/2}^{\left(p+1\right)}   \right)
.
\end{equation}

\noindent{}In the above relations, the Laplace coefficient is given by

\begin{equation}
b_{s}^{\left(m\right)} = \frac{1}{\pi} \int_{0}^{2\pi} \frac{\cos{\left(mx\right)} dx }
                       {\left(1 - 2 \alpha \cos{x} + \alpha^2 \right)^{s}}
\end{equation}

\noindent{}and satisfies the relation

\begin{equation}
\frac{d}{d \log{\alpha}} b_{s}^{(j)} = s \alpha \left( b_{s+1}^{\left(j-1\right)} - 2 \alpha b_{s+1}^{\left(j\right)} + b_{s+1}^{\left(j+1\right)}   \right)
.
\end{equation}

The conditions for capture (equations \ref{critn}-\ref{crite}) are sufficient, but not necessary. 
If capture is avoided, continuing migration towards the planet will allow the asteroid to encounter
an infinitely increasing number of $p+1$:$p$ resonances.  For a high enough value of $p$, the
gravitational reaches of these individual resonances cross one another, creating ``resonant overlap''.

\begin{figure*}
\includegraphics[angle=270,width=14cm]{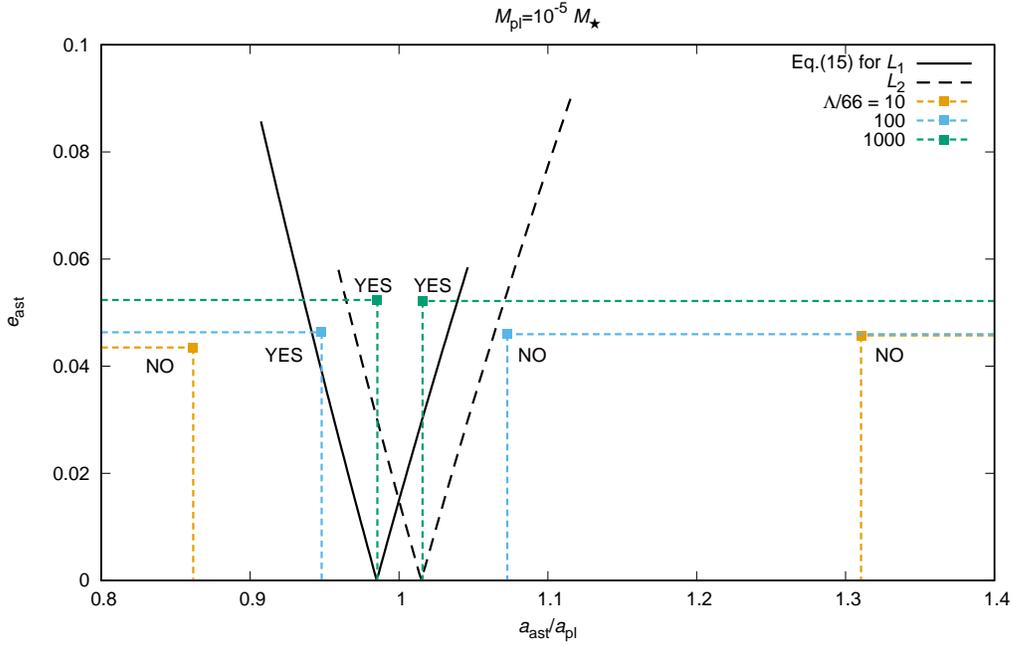}
\caption{
Semimajor axes ($x$-axis) and critical asteroid eccentricities ($y$-axis) at constant stellar luminosities (coloured dashed lines) for which a 100m-radius asteroid of density 3 g/cm$^3$ that is moving towards a planet (either inward or outward) could bypasses mean motion commensurabilities and instead be temporarily captured by the planet itself. These possibilities are illustrated by the ``YES'' and ``NO'' statements. The capture boundaries, given by the Lagrangian L1 and L2 regions, are shown by the black diagonal lines, and were derived in Higuchi \& Ida (2016, 2017). The orange, blue and green dashed lines correspond to asteroid speeds generated from an approximated Yarkovsky drift given by respectively, $10L_{\odot}$, $100L_{\odot}$, and $1000L_{\odot}$. This plot demonstrates that temporary capture may occur under the correct conditions, i.e. with the appropriate speed and quick dissipation mechanism. Yarkovsky drift which is enhanced from that in the solar system may provide this necessary speed.
}
\label{CaptEcc}
\end{figure*}

\begin{figure}
\includegraphics[angle=0,width=8cm]{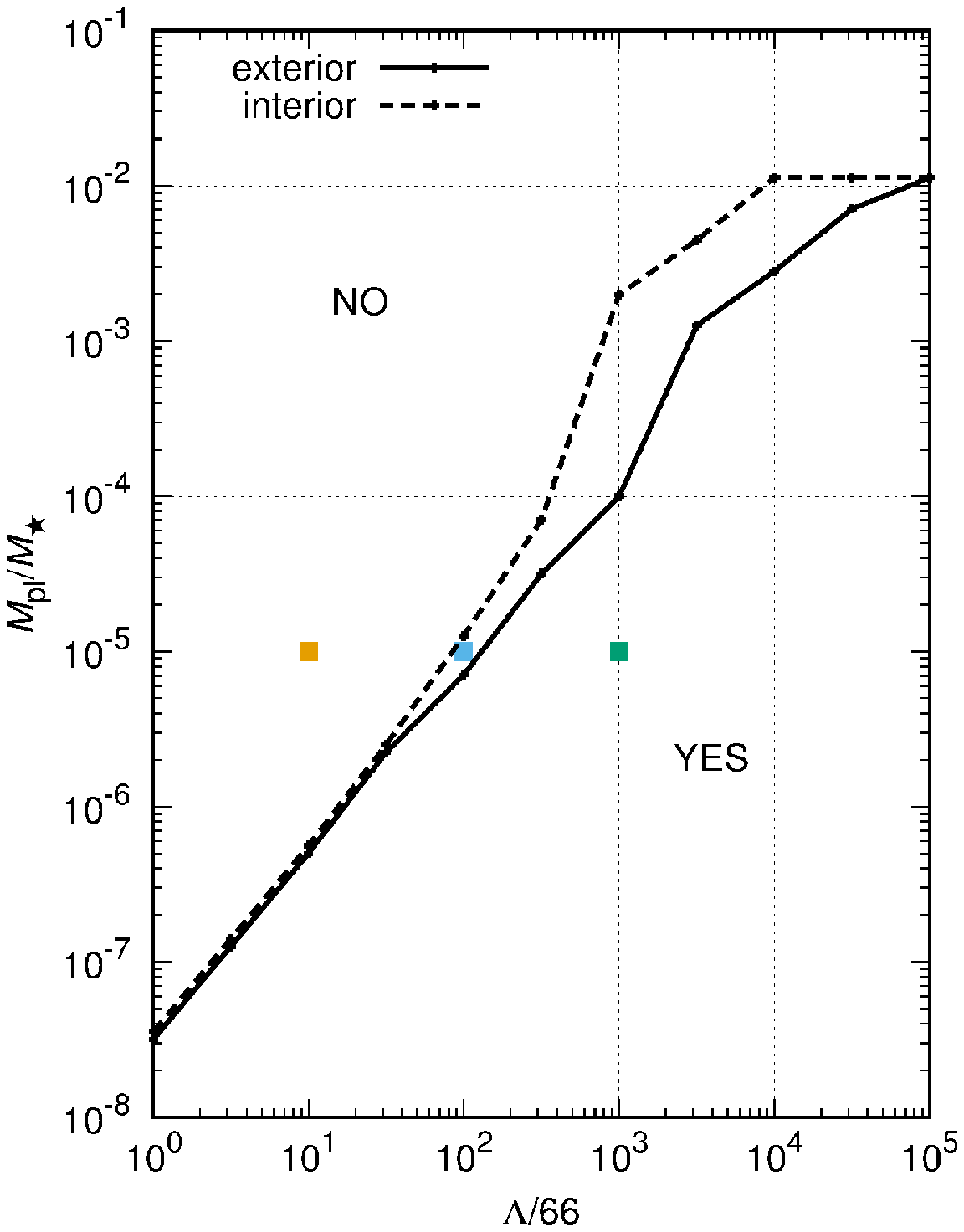}
\caption{
Like Fig. \ref{CaptEcc}, but as a function of stellar luminosity ($x$-axis, where $\Lambda = 66$ corresponds to
$1M_{\odot}$) and planetary mass ($y$-axis), for asteroids both interior (dashed line) and exterior (solid line) to the planet. The plot illustrates that {\forref the} current solar system-generated Yarkovsky {\forref effect} pushes asteroids too slowly to be captured around any major planet. However, if giant branch luminosity remained constant for sufficiently long, then capture would be possible around both terrestrial and giant planets.
}
\label{CaptMstar}
\end{figure}

\subsection{Capture into the planet's Hill sphere}

This resonant overlap can create gravitational instability, resulting in the asteroid's
escape from the system, or collision into the planet or star. The width of the 
overlap region has been estimated
analytically and numerically 
\citep{wisdom1980,quifab2006,muswya2012,bodqui2014,shaetal2015},
but is not yet known to admit an exact analytical solution.

If an asteroid is migrating towards a planet, avoids entrapment into a resonance during
migration, {\it and} passes through the resonant overlap region without being scattered away,
then the planet may capture the asteroid, at least temporarily. The conditions for capture 
have been quantified
by \cite{higida2016,higida2017}, who specify the combination of $a$ and $e$ values required for
temporary capture by a planet. Denote these values as $a_{\rm tc}$ and $e_{\rm tc}$. They
are given as curves on the $a-e$ plane that satisfy $q = a_{\rm tc} \left(1 \pm e_{\rm tc} \right) = r_{\rm pl} \pm r_{\rm Hill}$,
where $q$ is the periastron distance, $r_{\rm pl}$ is the heliocentric
distance of the planet, and 

\begin{equation}
r_{\rm Hill} = a_{\rm pl} \left(\frac{M_{\rm pl}}{3 M_{\star}}\right)^{1/3}
.
\end{equation}

First $a_{\rm tc}$ is computed through

\begin{equation}
a_{\rm tc} = a_{\rm pl} \left[\frac{2}{A} - \left(1 \pm \sqrt{3 \kappa} r_{\rm Hill} \right)^2 \right]^{-1}
\label{atc}
\end{equation}

\noindent{}where either $A = 1 - r_{\rm Hill}/a_{\rm pl}$, when $r_{\rm ast} < r_{\rm pl}$ at the L1 
point, or where $A = 1 + r_{\rm Hill}/a_{\rm pl}$, when $r_{\rm ast} > r_{\rm pl}$ at the L2 point. Also,
$0 \le \kappa \le 2$, which spans the allowed capture region.  Then, 
$e_{\rm tc}$ is computed through

\begin{equation}
e_{\rm tc} = 1 - \frac{a_{\rm pl} A}{a_{\rm tc}} \ \ {\rm or} \ \ e_{\rm tc} = \frac{a_{\rm pl} A}{a_{\rm tc}} - 1
.
\label{etc}
\end{equation}

\subsection{Link with giant branch Yarkovsky forces}

Now we link the resonant formalism above with the consequences of
differential migration due to Yarkovsky forces.

Drifts from Yarkovsky forces are unlikely to produce the type
of smooth migration that is often envisaged in a protoplanetary disc. The shape,
size and thermal properties of an asteroid determine the migration rate
\citep[e.g.][]{vokrouhlicky1998,broz2006,wanhou2017}, may change with time, and
are linked with YORP spin changes. Despite these complexities, a common
simplification for Yarkovsky-based studies of solar system asteroids is
to assume that only their semimajor axes change. This Solar-induced
drift for asteroids whose sizes are within the metre-to-kilometre range
is on the order of $10^{-5} - 10^{-2}$ au/Myr
\citep[e.g.][]{botetal2000,botetal2006,galetal2011}.

\subsubsection{Constant drift approximation}

We can now derive some results for the current solar system by using the well-established 
constant drift approximation. Doing so also helps lay the groundwork for later considering 
the giant branch phases of evolution.

\cite{veretal2015} showed in their Eq. (A2) that the averaged secular drift due to the Yarkovsky 
effect has the following dependency on semimajor axis

\begin{equation}
\frac{da_{\rm ast}}{dt} \propto \Lambda a_{\rm ast}^{-1/2}
\end{equation}

\noindent{}where $\Lambda$ is a factor given (or assumed) that is a function of asteroid density, size
and many other parameters. For an asteroid with a radius of $R_{\rm ast} = 100$ m 
and density of $3$~g/cm$^3$  in the current solar system (inhabited by a $1L_{\odot}$ Sun), 
we obtain $\Lambda \approx 66$ m$^{3/2}$/s. This value can be related to
a dimensionless constant $\Lambda_0 = 2.3 \times 10^{-7}$ where

\begin{equation}
\frac{da_{\rm ast}^{\rm exterior}}{dt} \approx \Lambda_0 \left(p+1\right)^{-1/3} p^{1/3} \sqrt{\frac{G M_{\ast}}{a_{\rm pl}}}
\ {\left[ \rm au \ ({\rm yr}/2\pi)^{-1}  \right]}
,
\label{dadtlamext}
\end{equation}

\begin{equation}
\frac{da_{\rm ast}^{\rm interior}}{dt} \approx \Lambda_0 \left(p+1\right)^{1/3} p^{-1/3} \sqrt{\frac{G M_{\ast}}{a_{\rm pl}}}
\ {\left[ \rm au \ ({\rm yr}/2\pi)^{-1}  \right]}
.
\label{dadtlamint}
\end{equation}

\noindent{}Scaling $\Lambda_0 = 2.3 \times 10^{-7}$ to different values allows us to probe different
magnitudes of the Yarkovsky effect easily.

The asteroid's migration is guaranteed to stop at the first 1st-order
resonance encounter after (and if) both $da/dt  < da_{\rm crit}/dt$ and $e<e_{\rm crit}$ are satisfied. In
order to find the value of $p_{\rm min}$ which is the minimum that satisfies both conditions, we set 
equations (\ref{dadtlamext}-\ref{dadtlamint}) equal to
equations (\ref{dadtcritext}-\ref{dadtcritint}), and then implicitly solve for $p = p_{\rm crit}$.
For external asteroids,

\begin{equation}
\Lambda_0 = \frac{5 \cdot 3^{2/3}}{4} \left(p_{\rm crit}+1\right)^{19/9}  p_{\rm crit}^{-16/9}
\left| \frac{M_{\rm pl}}{M_{\star}} \left(\alpha f_{31} + f_{\rm i} \right) \right|^{\frac{4}{3}}
.
\end{equation}

\noindent{}For internal asteroids,

\begin{equation}
\Lambda_0 = \frac{5 \cdot 3^{2/3}}{4} \left(p_{\rm crit}+1\right)^{-17/9}  p_{\rm crit}^{20/9}
\left| \frac{M_{\rm pl}}{M_{\star}} \left(f_{27} + f_{\rm e} \right) \right|^{\frac{4}{3}}
.
\end{equation}

Then $p_{\rm min}$ is the nearest integer greater than or equal to $p_{\rm crit}$
and $a_{\rm final}$ is given by equation (\ref{alphaeq})  as a function of $p_{\rm min}$ and $a_{\rm pl}$.
We can then obtain the critical eccentricity (equation \ref{crite}) by substituting
$p$ with $p_{\rm min}$.

Figures \ref{CaptEcc} and \ref{CaptMstar} quantify these analytical results. Figure \ref{CaptEcc} demonstrates
that for temporary Hill sphere capture to occur around a $10^{-5} M_{\star}$ planet, the Sun's current luminosity
would need to be increased by at least a factor of 100. An asteroid with a density of 3 g/cm$^3$ and a radius of
$R_{\rm ast} = 100$ m would also need to maintain an eccentricity of under about 0.06 throughout its migration.
Hill sphere capture could occur from either internal or external migration. Figure \ref{CaptMstar} then quantifies
this migration boundary as a function of both planetary mass and stellar luminosity, and illustrates that temporary
Hill sphere capture may occur around any type of planet provided that the stellar luminosity is sufficiently high.
Such capture, however, requires orders-of-magnitude higher luminosities for giant planets than for terrestrial planets.

\subsubsection{Variable drift approximation}

However, applying a constant drift in only the semimajor axis throughout the giant branch
phases of evolution would be unrealistic. The stellar luminosity varies non-monotonically
and by three orders of magnitude while the asteroid's orbit is expanding anyway through
stellar mass loss. Further, the asteroid's eccentricity and inclination do not remain
constant, the former potentially changing by up to 0.08/Myr \citep{veretal2015}. Therefore,
whether asteroid eccentricities can remain under the critical eccentricity that is
given by equation (\ref{crite}) is unclear, as is the consequences of inclination evolution. 
Although at the beginning
of the red giant and asymptotic giant branch phases the luminosity is low enough
to potentially achieve resonant capture, the chances of capture would likely
decrease towards the end of each subphase.

Modelling these complexities remains a challenge. In the forthcoming section, we pursue
an extension to previous work, but one simple enough to be computationally feasible.

\section{Simulating Yarkovsky drift}

Having now detailed the potential limitations of the analogy with
planetary migration in a disc, we
set about placing bounds on Yarkovsky-induced movement of an asteroid in the vicinity of a planet during the giant branch phase.

Asteroids are orbitally perturbed from stellar radiation which is (i) absorbed,
(ii) immediately reflected, and (iii) reflected 
after a delay. As described in detail by \cite{veretal2015}, the first two
processes give rise to what is known as {\it Poynting-Robertson drag} and {\it radiation pressure}.
The third component gives rise to the Yarkovsky effect, which ``turns on'' only for objects larger than
about 1-10 m in size.

Crucially, \cite{veretal2015} showed that acceleration due to Yarkovsky effect is proportional 
to $(1/c)$, whereas acceleration due
to Poynting-Robertson drag and radiation pressure is proportional to $(1/c^2)$, where $c$ is the speed
of light. Consequently, contributions from Poynting-Robertson drag and radiation pressure are negligible
in this context.

\subsection{The true Yarkovsky acceleration}

The acceleration due to the Yarkovsky effect is (equation 27 of \citealt*{veretal2015})

\begin{equation}
\left(\frac{d \vec{v}_{\rm ast}}{dt}\right)_{\rm Yar}
=
\frac{A_{\rm ast}L_{\star}(t)}{4\pi M_{\rm ast} c r_{\rm ast}^2}
\left[Q_{\rm PR} \mathbb{I} +  k Q_{\rm Yar} \mathbb{Y}(t)  \right]\vec{\iota}
\label{yarfin}
\end{equation}


\noindent{}where $\vec{v}_{\rm ast}$ is the velocity of the asteroid, 
$k$ is a constant between 0 and 1/4 that is strongly 
linked to the asteroid's rotational state, 
$A_{\rm ast}$ is the momentum-carrying 
cross-sectional area of the asteroid's heated surface, $L_{\star}(t)$ is the stellar luminosity
(with an emphasis on time dependence), $Q_{\rm PR}$ is a constant that equals 
the sum of the target's absorption
efficiency and reflecting efficiency,
and $Q_{\rm Yar}$ is a constant between 0 and 1 that refers to the
difference between the asteroid's absorption efficiency and albedo. The matrix
$\mathbb{I}$ is the identity matrix, and $\mathbb{Y}$ is the Yarkovsky matrix, where
the absolute value of each entry is less than or equal to unity.
The vector $\iota$ is the relativistic direction correction:

\begin{equation}
\vec{\iota}
\equiv
\left(1 - \frac{\vec{v}_{\rm ast} \cdot \vec{r}_{\rm ast}}
{cr_{\rm ast}} \right) \frac{\vec{r}_{\rm ast}}{r_{\rm ast}}
- \frac{\vec{v}_{\rm ast}}{c},
\label{PRBurns2}
\end{equation}

The greatest complexity in equation (\ref{yarfin}) arises from the position-, 
velocity- and time-dependent 3-by-3
matrix $\mathbb{Y}$, which is a function of the asteroid's spin axis, specific angular momentum,
emissivity, specific heat capacity and thermal conductivity. The particular functional forms and resulting
matrices are in turn dependent on the adopted heat conduction model. The spin evolution
is further dictated
by the YORP effect, which is a function of the asteroid's shape.

\subsection{A simplified Yarkovsky acceleration}

Modelling all of these complications is well beyond the scope of this study. Instead,
we pursue a simpler approach, but one which is still more detailed than any previous
giant branch Yarkovsky study \citep{veretal2015}: here we place limits on the resultant
motion by adopting different constant entries for $\mathbb{Y}$. This matrix contains
all of the Yarkovsky physics, and by setting its entries to their extreme values, we can
bound possible asteroid motions.

Therefore, we set $Q_{\rm PR} = Q_{\rm Yar} = 1$ and  $k = 1/4$, and assume 
that the entries for $\mathbb{Y}$ are constant throughout the simulation.
We also assume the asteroid is a perfect sphere with radius $R_{\rm ast}$ 
(thereby preventing initiation of the YORP effect), such that

\begin{equation}
\left(\frac{d \vec{v}_{\rm ast}}{dt}\right)_{\rm Yar}
=
\frac{R_{\rm ast}^2 L_{\star}(t)}{4 M_{\rm ast} c r_{\rm ast}^2}
\left[\mathbb{I} +  \frac{1}{4} \mathbb{Y}  \right]
\vec{\iota}
.
\label{yarfi2}
\end{equation}

\noindent{}The functional form in equation (\ref{yarfi2}) highlights
the strong dependence of Yarkovsky acceleration on asteroid radius.

\subsubsection{Secular trends}

Before choosing values of $\mathbb{Y}$ and running simulations, we can obtain some analytic 
results based on the simplified acceleration in equation (\ref{yarfi2}).
\cite{veretal2015} analytically integrated a similar acceleration,
and their appendix presents the resulting secular evolution of orbital elements\footnote{In each of their
Eqs. A4-A5, the first dot refers to matrix multiplication, and the second dot refers to dot product. In this
context, for those equations the transpose of the $\mathbb{Q}$ should be applied instead of the traditional form
of $\mathbb{Q}$.}.
In order to relate our results to theirs, we set

\begin{equation}
\mathbb{Q} \equiv  \mathbb{I} +  \frac{1}{4} \mathbb{Y}
.
\end{equation}

Then, we can observe several important trends, which will be helpful for understanding our
results:

\begin{itemize}

\item The Yarkovsky-induced semimajor axis evolution $\left \langle da_{\rm ast}/dt \right \rangle$ is completely independent
of $\mathbb{Q}_{11}, \mathbb{Q}_{22}$ and $\mathbb{Q}_{33}$.

\

\item The Yarkovsky-induced eccentricity evolution $\left \langle de_{\rm ast}/dt \right \rangle$ does not tend towards infinity
for any value of $e_{\rm ast}$.  

\

\item In the limit of circular, coplanar asteroid orbits,

\begin{equation}
\left \langle \frac{da_{\rm ast}}{dt} \right \rangle \propto \frac{\mathbb{Q}_{21} - \mathbb{Q}_{12}}{\sqrt{a_{\rm ast}} \left(1 -  e_{\rm ast}^2 \right)}
,
\label{aevo}
\end{equation}

\begin{equation}
\left \langle \frac{de_{\rm ast}}{dt} \right \rangle \rightarrow 0
,
\label{eevo}
\end{equation}


\

\item For non-coplanar, near-circular orbits where the $\mathbb{Q}_{21}$ and/or $\mathbb{Q}_{12}$ terms dominate, as in equation (\ref{aevo}), then

\begin{equation}
\left \langle \frac{di_{\rm ast}}{dt} \right \rangle \propto \frac{
\sin{i}\left(
\mathbb{Q}_{12} \sin^2{\Omega_{\rm ast}} -  \mathbb{Q}_{21} \cos^2{\Omega_{\rm ast}}  
\right)
}{a_{\rm ast}^{3/2}}
,
\label{ievo}
\end{equation}

\end{itemize}

\noindent{}where $\Omega_{\rm ast}$ represents the longitude of ascending node of the asteroid. 
These trends will help set initial conditions for our simulations and explain our results 
as we sample limiting cases of motion.

\subsubsection{Equations of motion}

The Yarkovsky-induced acceleration on the asteroid must be combined with the
acceleration arising from stellar mass loss and the interaction with the planet.
The planet is also accelerating due to the mass loss, but is negligibly perturbed
by Yarkovsky acceleration due to its size.

As explained by \cite{hadjidemetriou1963} and \cite{veretal2013a}, this acceleration
from mass loss is automatically taken into account through the equations of motion 
through the
change of stellar mass. The final equations of motion, assuming that the asteroid 
does not perturb the planet, are

\begin{equation}
  \frac{d^2\vec{r}_{\rm pl}}{dt^2} = -G M_{\star}(t)\frac{\vec{r}_{\rm pl}}{\left|\vec{r}_{\rm pl}\right|^3}
  ,
\label{planevo}
\end{equation}

\[
\frac{d^2\vec{r}_{\rm ast}}{dt^2} = -G M_{\star}(t)\frac{\vec{r}_{\rm ast}}{\left|\vec{r}_{\rm ast}\right|^3} 
                          -G M_{\rm pl} \frac{\vec{r}_{\rm ast} - \vec{r}_{\rm pl}}{\left|\vec{r}_{\rm ast} - \vec{r}_{\rm pl}\right|^3}
\]
 
\begin{equation}
\ \ \ \ \ \ \ \ \ \ \ \ \
+ 
\frac{R_{\rm ast}^2 L_{\star}(t)}{4 M_{\rm ast} c r_{\rm ast}^2}
\mathbb{Q}\vec{\iota}
.
\end{equation}

We have chosen to fix the mass of the planet, which may not be a suitable approximation for planets which are primarily composed of an atmosphere and reside close enough to the giant star to undergo partial or complete evaporation 
\citep{livsok1984,neltau1998,soker1998,villiv2007,wicetal2010}. Determining different evaporation prescriptions due to planet core, mantle and atmosphere compositions, as well as the type of flux emitted from the star, is beyond the scope of this study (but represents an important future consideration).

\subsubsection{Integrator}

In order to integrate the equations of motion, we inserted the simplified 
Yarkovsky acceleration (equation \ref{yarfi2})
into the combined stellar and planetary evolution code that was
first used in \cite{musetal2018}. This code is a modification
of the code established in \cite{veretal2013a}, and is originally based on the {\tt Mercury} integration
suite \citep{chambers1999}. The code utilises the RADAU integrator to dynamically evolve the system, 
and interpolates stellar mass and radius at subdivisions of each timestep. The stellar profiles of mass, radius and luminosity are obtained from the {\it SSE} stellar
evolution code \citep{huretal2000}. 

There is a balance between integrator speed and accuracy. Because we modelled the possibility of
entrapment into resonance, we required a high-enough accuracy to track orbital angles to within a few
degrees over a few Myr. Figure A1 of \cite{musetal2018} suggests that an accuracy of $10^{-11}$ suits our purposes, a value that we adopted.

\subsection{Initial conditions}


Our limited computational resources prevented us from performing a broad sweep of phase space.
We instead focussed on the most relevant cases, and considered 

\begin{itemize}

\item Only the asymptotic giant branch phase of an initially $2M_{\odot}$ star. This phase lasts for about 1.71 Myr, and our integrations ran for up to 2 Myr from the start of that stellar phase. At that initial time, the stellar mass had already been reduced only by about $1.2 \times 10^{-4} M_{\odot}$. Throughout this stellar phase, the star lost about 68 per cent of its mass in route to becoming a white dwarf. This mass loss corresponds to an adiabatic orbital semimajor axis increase of a factor of about 3.1.

\item A Neptune-mass planet located at 30 au from the host star at the beginning of the simulation.  Its eccentricity and inclination were set initially at exactly zero to provide a reference for the asteroid evolution. {\rev We chose a Neptune-mass planet because exo-Kuiper belt objects are thought to be the primary source of white dwarf pollution. Hence, a rough solar system analogue was appropriate as a first guess. Other investigations have shown, however, that lower-mass planets are more efficient polluters \citep{frehan2014,musetal2018}, whereas Fig. 2 of this paper reveals that higher-mass planets are more efficient resonant trappers.}

\item Asteroids in belts both internal and external to the planet. In all simulations, these asteroids were given initial semimajor axes selected from a random uniform distribution ranging from both 10-25 au and 35-50 au. The inner boundary of 10 au is far enough away from the star to ensure that at least some asteroids at those locations would survive YORP-induced rotational fission \citep{veretal2014a}, which we do not model here.

\item 200 asteroids per simulation. The asteroids were given a density of 3 g/cm$^3$ and radii ranging from 100~m to 1000~km. These asteroids were treated as ``small'' particles in our integrator, a simulation designation which indicates that the asteroids do not perturb each other although they do perturb the planet (thereby introducing additional, but negligible terms in equation \ref{planevo}).

\item Asteroids which were initialised at the beginning of the simulation with inclinations chosen from a random uniform distribution ranging {\forref from $0^{\circ}$} to $10^{\circ}$.

\item Asteroids which were initialised at the beginning of the simulation with two different eccentricity distributions depending on the simulation. The first is circular asteroids, and the second are asteroids with eccentricities chosen between $0.3$ to $0.7$. We emphasise that we are not attempting to reproduce realistic exo-debris discs in these systems (which would require a more detailed Yarkovsky model), but rather are placing bounds and observing trends. {\rev The evolution of most initially eccentric asteroids that we simulated are not shown, unless they provide a revealing physical trait.}

\item Asteroids which were initialised at the beginning of the simulation with orbital angles chosen from a random uniform distribution across their entire ranges.


\item Yarkovsky prescriptions according to the following values of $\mathbb{Q}$:

\begin{equation}
{\rm Model \ A:} \
\mathbb{Q} = 
\left( \begin{array}{ccc} 
            1 &  0 & 0  \\[3pt]
             \frac{1}{4} &  1 &  0 \\[3pt]
            0 &  0 &  1 \\
            \end{array}
            \right)
,
\end{equation}

\begin{equation}
{\rm Model \ B:} \
\mathbb{Q} = 
\left( \begin{array}{ccc} 
            1 &  \frac{1}{4} & 0  \\[3pt]
            0 &  1 &  0 \\[3pt]
            0 &  0 &  1 \\
            \end{array}
            \right)
,
\end{equation}

\begin{equation}
{\rm Model \ C:} \
\mathbb{Q} = 
\left( \begin{array}{ccc} 
            1 &  \frac{1}{4} & 0  \\[3pt]
            \frac{1}{4} &  1 &  0 \\[3pt]
            0 &  0 &  1 \\
            \end{array}
            \right)
.
\end{equation}



\end{itemize}

\noindent{}Our choices for Models A-C arise from our intention to place bounds on the motion
by considering only extreme cases. We approximate these extreme cases by focusing in on $\mathbb{Q}_{12}$
and $\mathbb{Q}_{21}$ from the analytic limits in equation (\ref{aevo}). 
{\rev In
terms of semimajor axis, Model A pushes asteroids outward, Model B pushes asteroids inward,
and Model C does not push them outward nor inward. Alternatively, Models A and B have only a minor effect
on asteroid inclination evolution, whereas Model C has a relatively large effect. }
We actually ran additional
simulations with other combinations of $\mathbb{Q}$ entries, but found that they represent intermediate
cases to Models A-C, supporting the analytics and our choices.




\begin{figure}
\includegraphics[width=8.8cm]{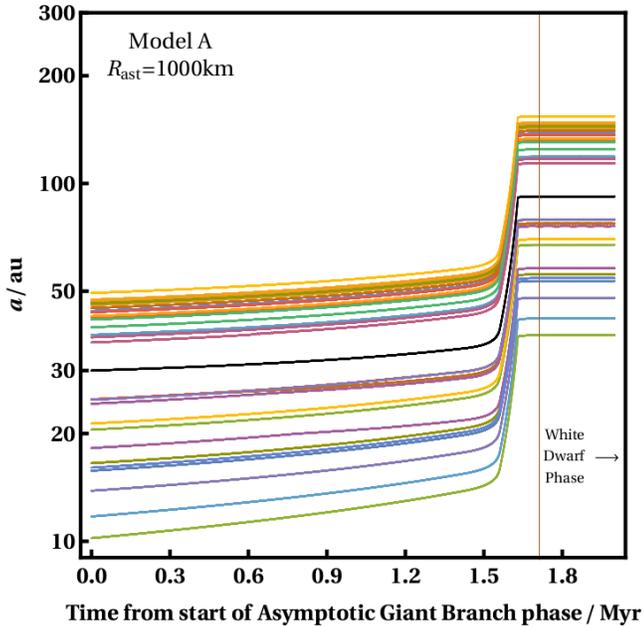}
\caption{
The semimajor axis evolution of a Neptune-mass planet (black line), an inner asteroid disc and an outer asteroid disc during the asymptotic giant
branch phase of an initially $2M_{\odot}$ (A-type) star. The thin vertical line indicates the time at which the star becomes a white dwarf.
Here the Yarkovsky Model A is applied, which
represents an extreme (bounding) case that would increase semimajor axis drift by the largest possible
extent. However, because the asteroids have $R_{\rm ast}=10^3$ km, they are large enough to be negligibly affected by {\forref the}
Yarkovsky {\forref effect}; the semimajor axis increase in this plot then arises purely from stellar mass loss, providing a convenient benchmark for comparison. Further, the planet-asteroid
interactions are minor enough to not be discernible on the plot.
}
\label{Test1410}
\end{figure}

\begin{figure}
\includegraphics[width=8.8cm]{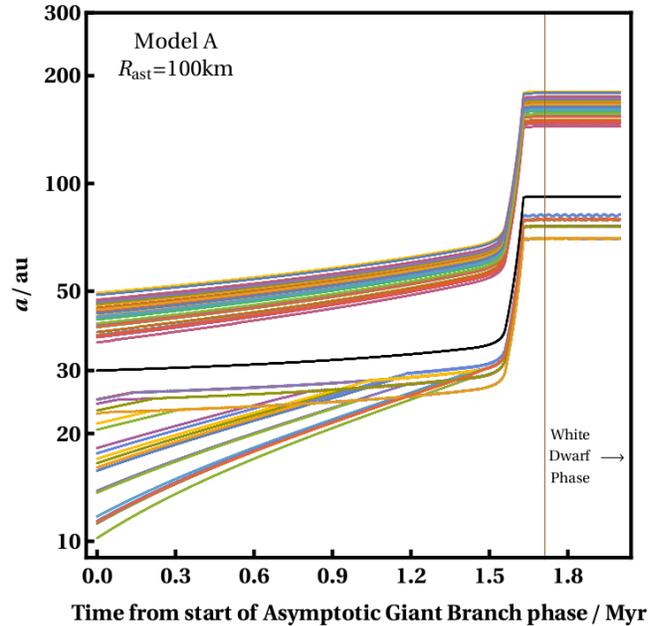}
\caption{
Like Fig. \ref{Test1410}, except here $R_{\rm ast}=100$ km. The Model A Yarkovsky effect on asteroids
is now noticeable, and increases the semimajor axes of the asteroids to a larger extent than in
Fig. \ref{Test1410}. {\rev The Yarkovsky effect can also be detected by comparing the bottommost curves
in Fig. \ref{Test1410} with those here: the Yarkovsky effect is maximised when $a_{\rm ast}$ is minimised,
and hence produces the steepest curves. Kinks in the curves for the asteroids in the inner planet indicate locations where resonant capture occurs.}
}
\label{Test1310}
\end{figure}

\begin{figure}
\includegraphics[width=8.8cm]{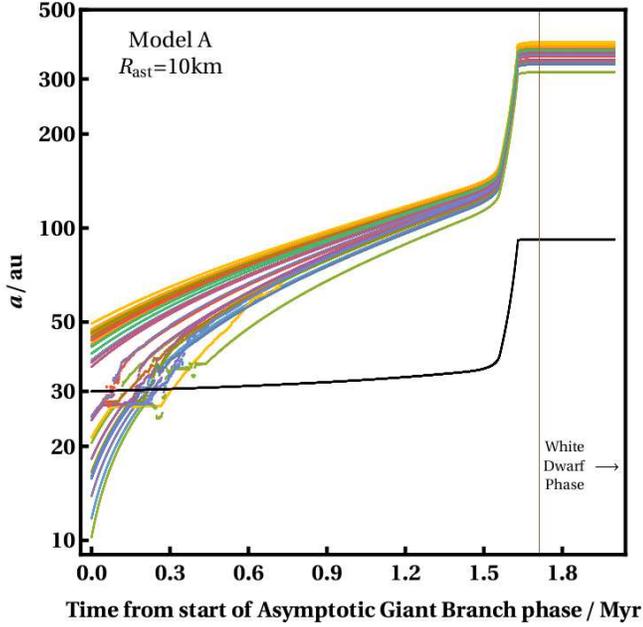}
\includegraphics[width=8.8cm]{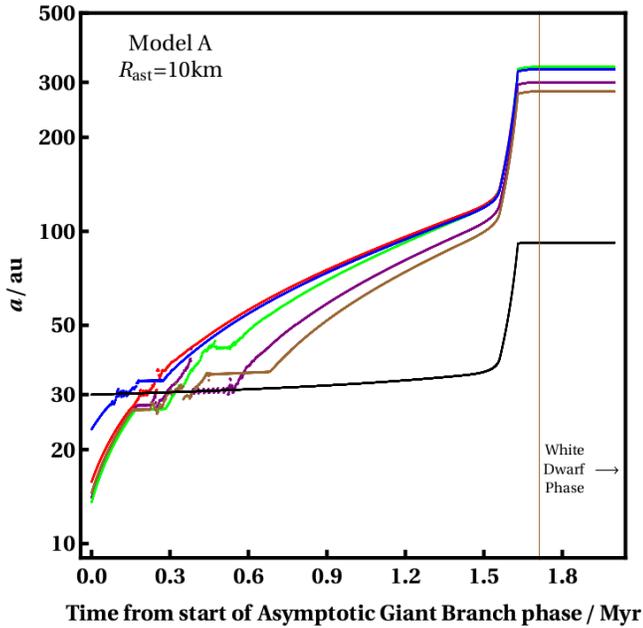}
\caption{
{\rev
Like Fig. \ref{Test1410}, except here $R_{\rm ast}=10$ km. Asteroids of this
size cross the planet's orbit due to the Yarkovsky effect. The
bottom panel highlights five asteroids from the top panel which are temporarily trapped in
resonance with the planet (black line) before being torn away due to the Yarkovsky
effect, and evolving independently of the planet thereafter. 
}
}
\label{Test1210}
\end{figure}

\begin{figure}
\includegraphics[width=8.8cm]{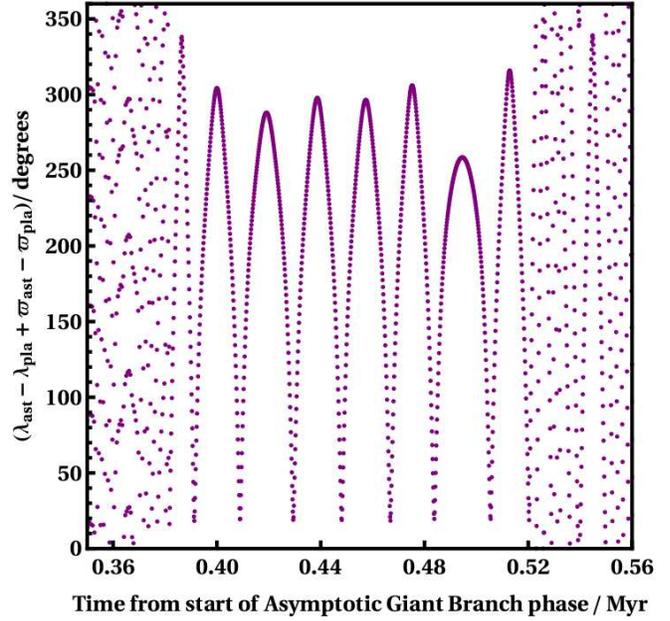}
\includegraphics[width=8.8cm]{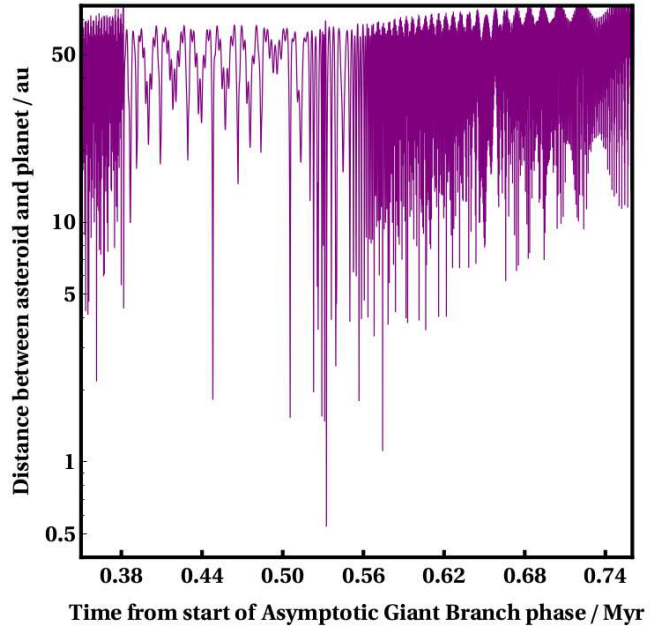}
\caption{
{\rev
Dynamical evolution of the purple asteroid in the lower panel of  Fig. \ref{Test1210}, whose semimajor
axis oscillates about the planet for the longest time (for 0.14 Myr, starting at 0.38 Myr). The upper 
panel illustrates the libration
of a $1$:$1$ resonant angle, proving that the asteroid is temporarily trapped in a co-orbital resonance.
The lower panel illustrates that despite this co-orbital motion, the asteroid is not trapped within the 
planet's Hill sphere, despite puncturing it occasionally.
}
}
\label{Test1210res}
\end{figure}

\begin{figure}
\includegraphics[width=8.8cm]{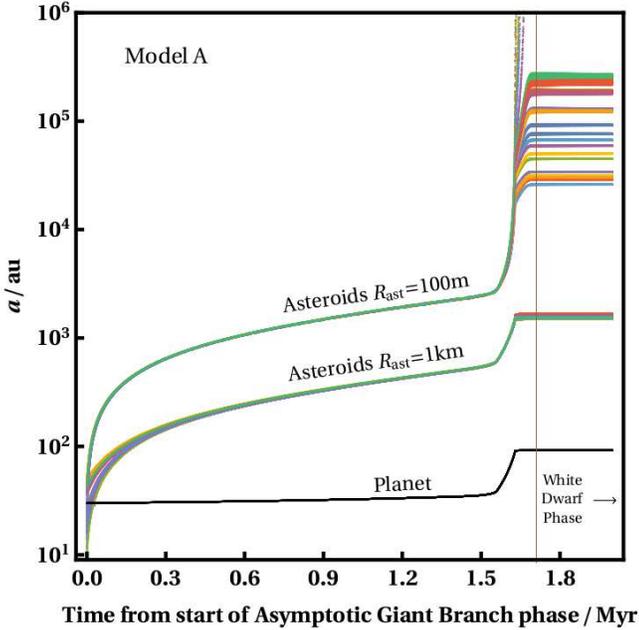}
\caption{
Like Figs. \ref{Test1410} and \ref{Test1310}, but now for {\rev two} smaller
asteroid sizes ($R_{\rm ast}=1$ km and $100$ m). Shown is the
superposition of the two different simulations with those asteroid sizes.
The semimajor axis increase for 100 m asteroids is so great that they 
can easily leave the Hill ellipsoid of the star, or be placed in a weakly
bound state subject to Galactic tides and stellar flybys.
}
\label{Comboa}
\end{figure}

\begin{figure*}
\centerline{
\includegraphics[width=8.8cm]{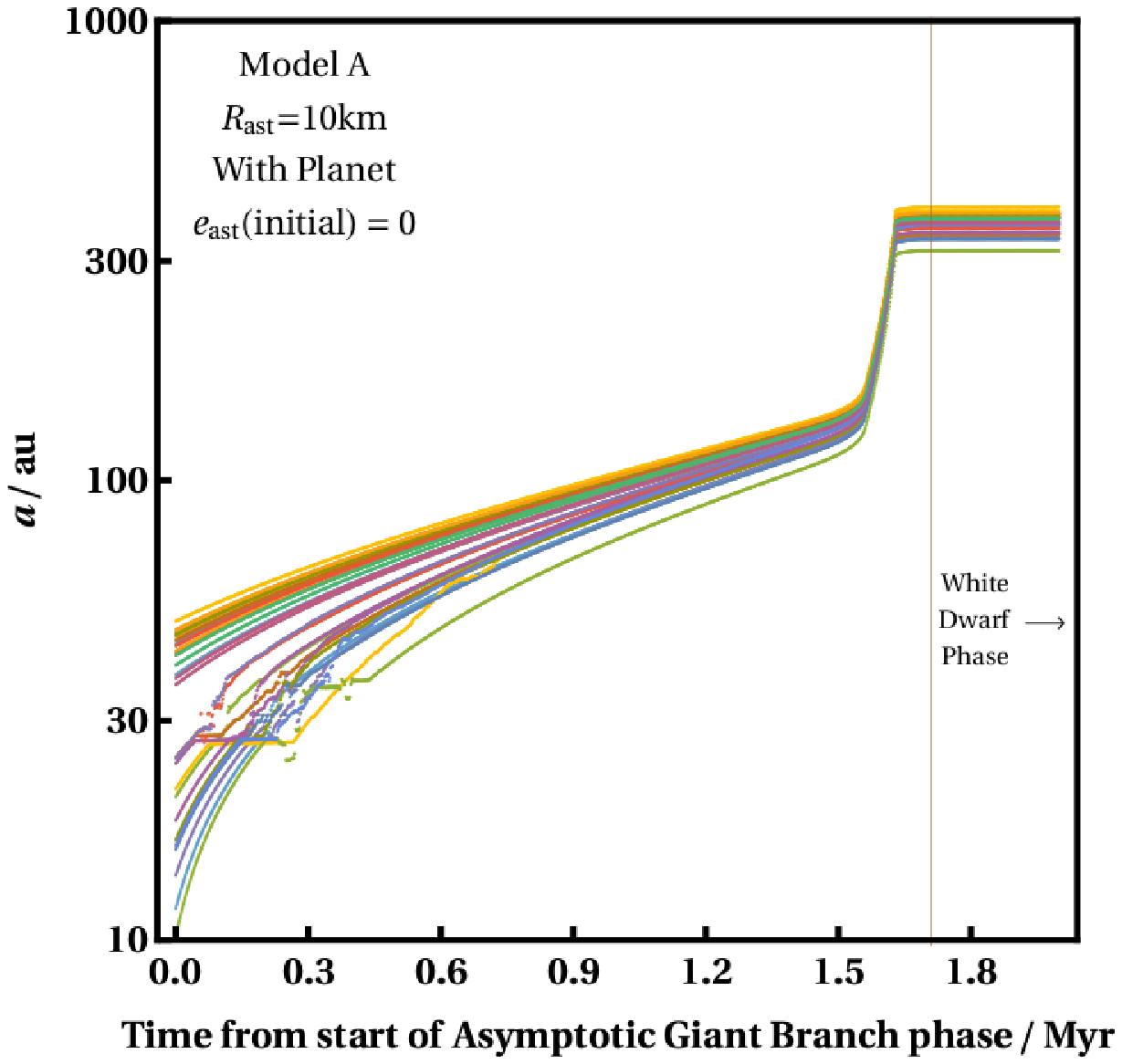}
\ \ \ \ \ \ \ \
\includegraphics[width=8.8cm]{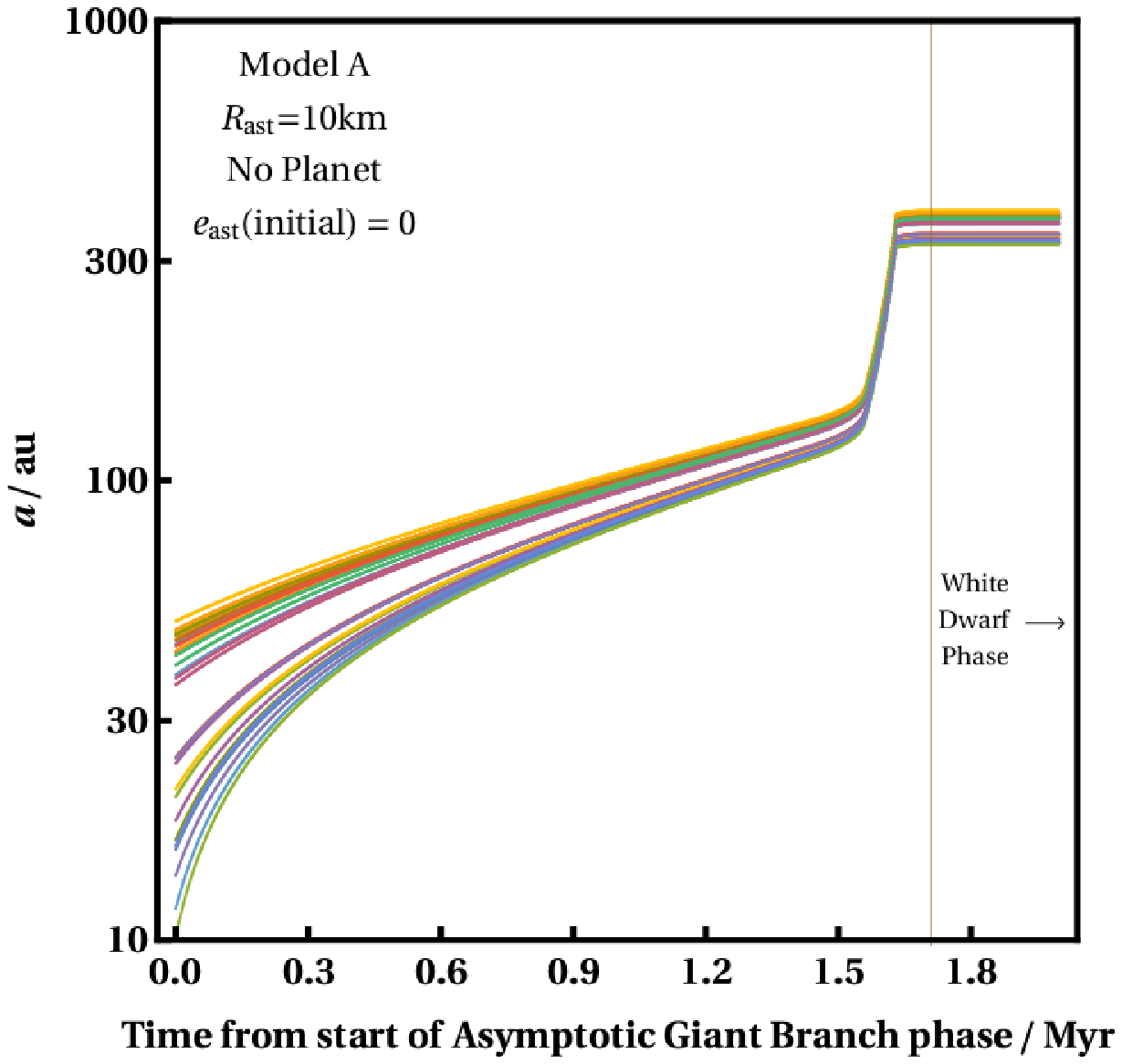}
}
\centerline{
\includegraphics[width=8.8cm]{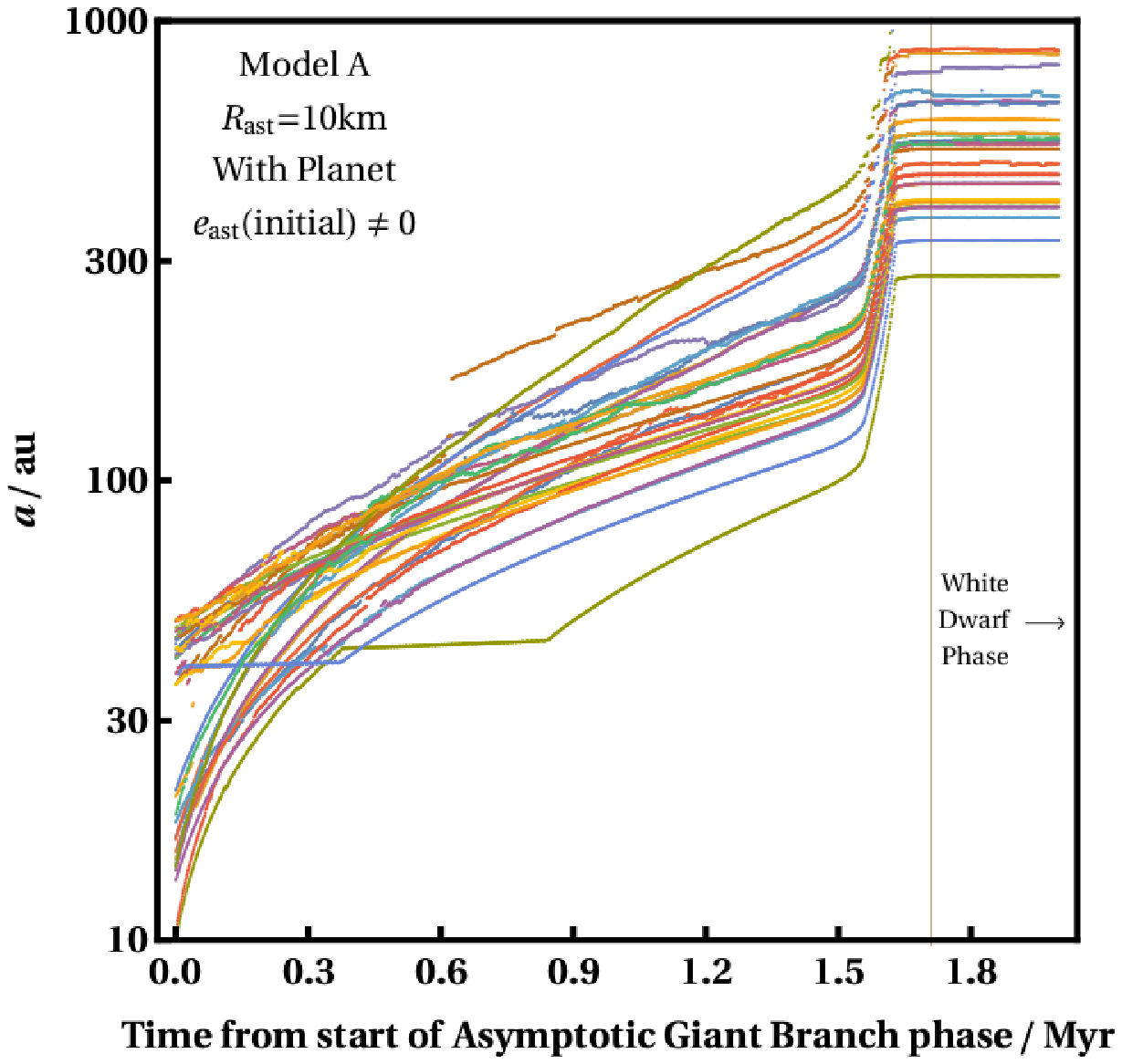}
\ \ \ \ \ \ \ \
\includegraphics[width=8.8cm]{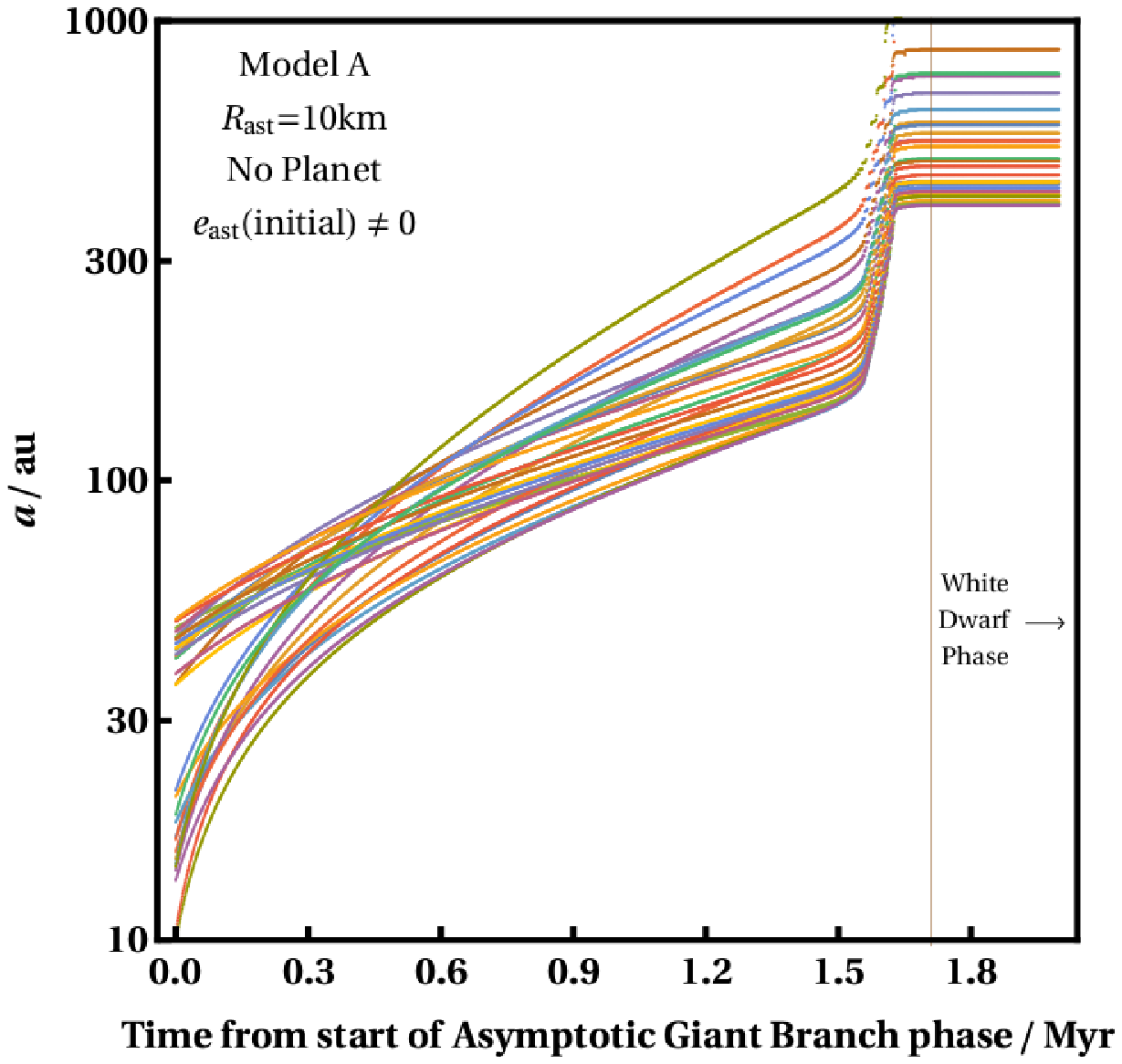}
}
\caption{
The influence of the planet (left panels) and initially eccentric orbits (lower panels) on the semimajor axis evolution 
of the $R_{\rm ast} = 10$ km asteroids which appear in {\rev the upper panel} of Fig. \ref{Test1210}. Both the planet
and the initial asteroid eccentricity change the final spread of semimajor axes that are attained along the white dwarf phase.
However, qualitatively, the evolution is similar. 
}
\label{Comp1210a}
\end{figure*}

\begin{figure*}
\centerline{
\includegraphics[width=8.8cm]{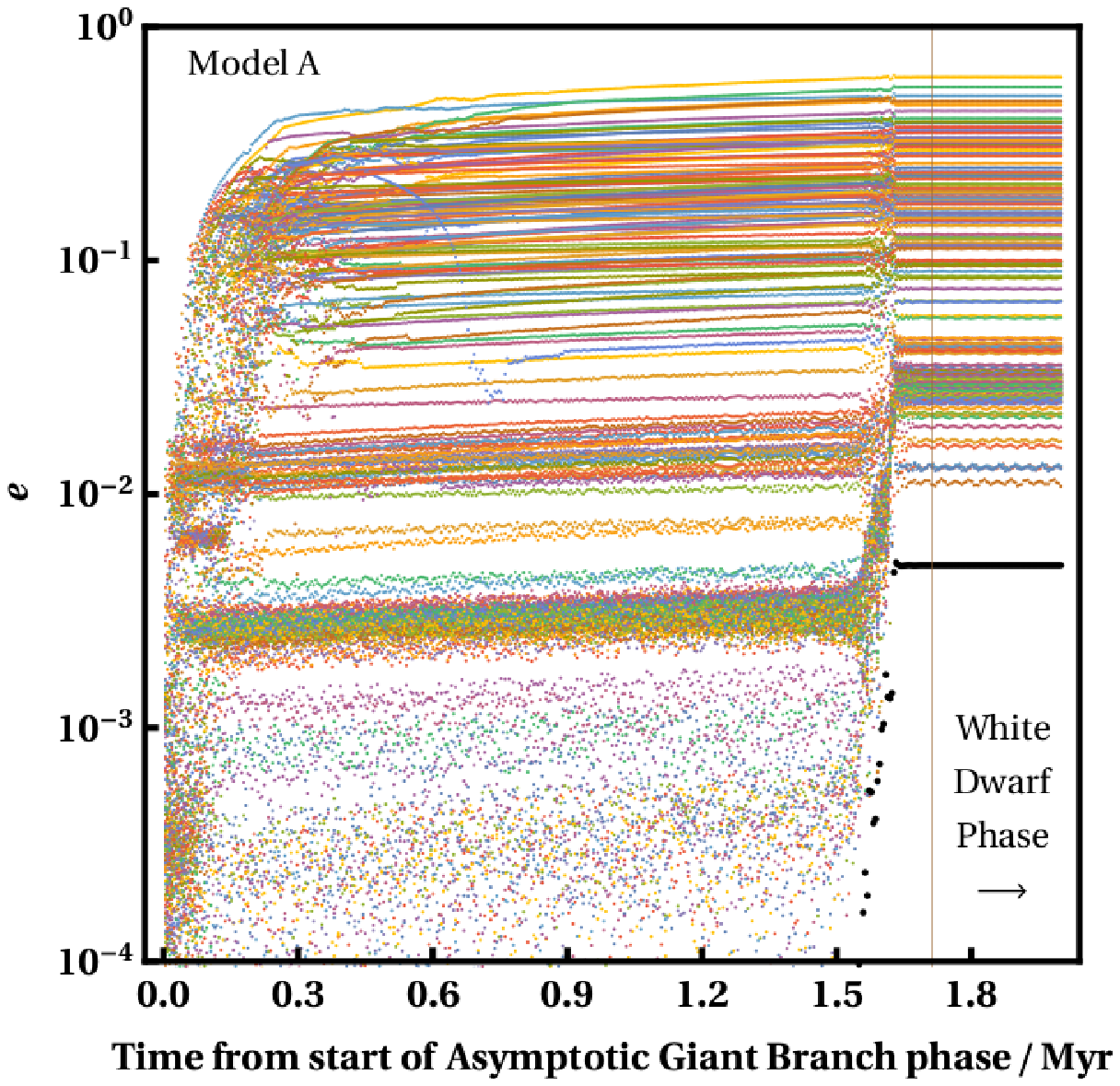}
\ \ \ \ \ \ \ \
\includegraphics[width=8.8cm]{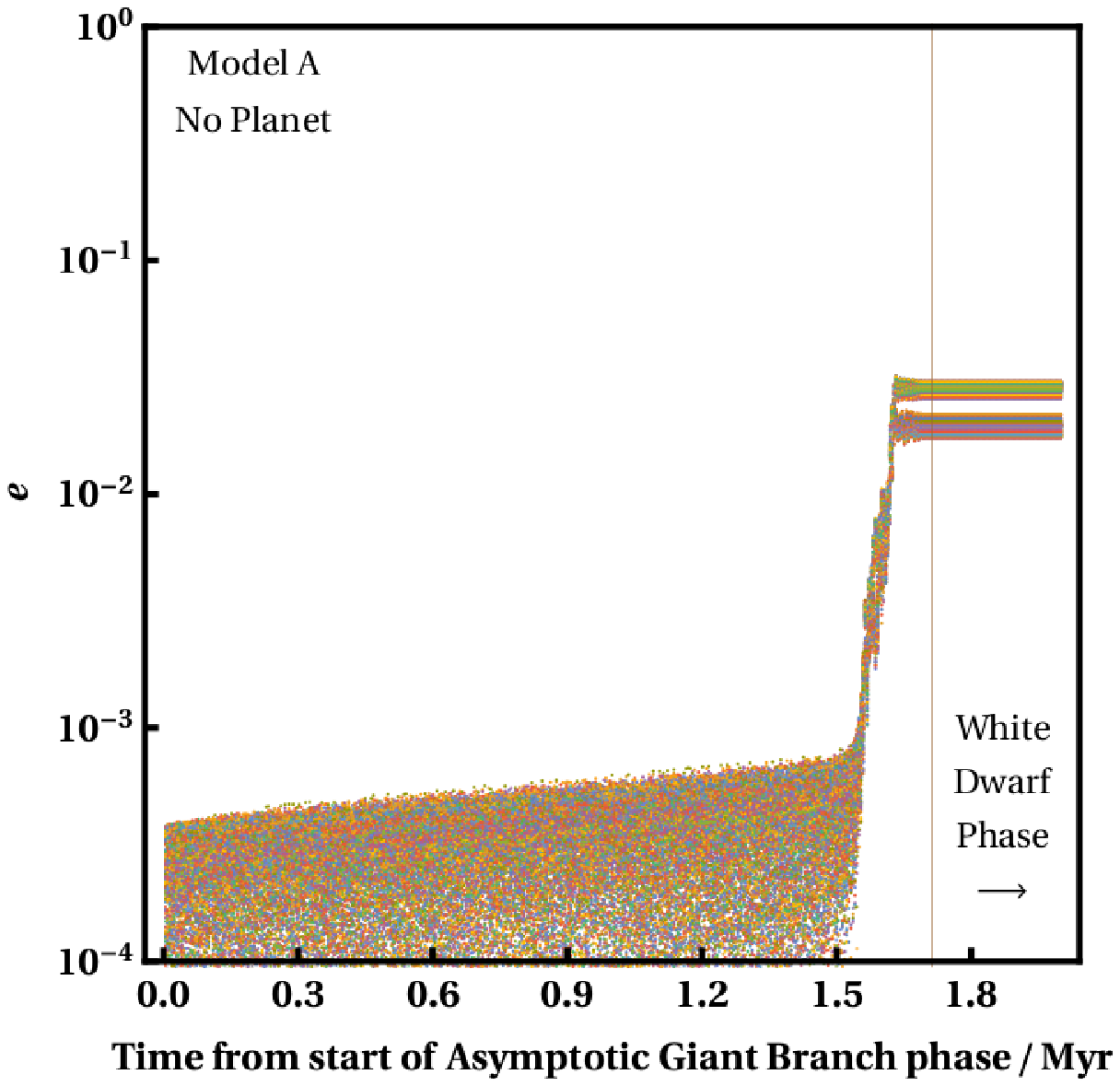}
}
\caption{
The influence of the planet on the eccentricity evolution of the asteroids which appear in the upper panels of Fig. \ref{Comp1210a} due to {\forref the} Model A Yarkovsky {\forref effect}.
When no planet is present (right panel), the eccentricity evolution is due entirely to stellar mass loss.
The planet, otherwise, dominates the eccentricity evolution (left panel), and does so immediately (within $10^5$ yr). This jump in eccentricity may explain why resonant capture is so infrequent (equation \ref{crite}).
}
\label{Comp1210e}
\end{figure*}

\section{Simulation results}

The asteroids in our simulations are accelerated by three major effects: (i) mass loss from the star, (ii) perturbations from the planet, and the (iii) {\forref radiation-driven} Yarkovsky {\forref effect}\footnote{Recall that Poynting-Robertson drag is weaker than {\forref the} Yarkovsky {\forref effect} by a factor of $\left( 1/c \right)$ and is anyway included our formalism through $\mathbb{Q}_{11}$, $\mathbb{Q}_{22}$, and $\mathbb{Q}_{33}$.}. Mass loss always pushes the asteroids away from the star (the orbital pericentre is always increasing, even in nonadiabatic motion; \citealt*{veretal2011}), whereas the planet and {\forref the} Yarkovsky {\forref effect} may perturb the asteroids in either direction. Although both mass loss and planets can increase $e_{\rm ast}$ to unity (causing escape; e.g. \citealt*{veretal2011,adaetal2013}), Yarkovsky acceleration cannot. \cite{veretal2015} investigated the relative magnitudes of stellar mass loss and {\forref the} Yarkovsky {\forref effect}, but did not include a planet nor run numerical simulations.

The Yarkovsky drift varies in a complex manner, and is a function of all orbital elements, even in our simple approximation where the matrix elements of $\mathbb{Q}$ remain constant over time. However, equations (\ref{aevo})-(\ref{eevo}) reveal that for small inclinations, the key parameters are $\mathbb{Q}_{12}$ and $\mathbb{Q}_{21}$. Although the Yarkovsky acceleration drops roughly as $1/\sqrt{a_{\rm pl}}$, the acceleration increases linearly as a function of $L_{\star}(t)$. The time dependence of $L_{\star}(t)$ is, in turn, nonlinear, and dominates changes in the Yarkovsky acceleration. Mass loss is largely adiabatic within about $10^3$ au \citep{veretal2011}, meaning that within this region it feeds back into the Yarkovsky acceleration only through an increase in asteroid semimajor axis.

Although each simulation contained 200 asteroids, we display the evolution of {\rev subsets of these on several of the figures} for better clarity and reduced filesizes. {\rev Nevertheless, we have looked at the evolution of all 200 asteroids in each case to identify any noteworthy behaviour.}

\subsection{Model A}

\subsubsection{Semimajor axis evolution}

For Model A, $\mathbb{Q}$ was chosen to maximize the asteroid's outward drift due the Yarkovsky effect. This drift is independent of the diagonal terms (Eq. A2 of \citealt*{veretal2015}), and is outward from equation (\ref{aevo}).  In Figs. \ref{Test1410}-\ref{Comboa}, we illustrate the extent of this drift for asteroids ranging in radius over five orders of magnitude. In all these figures, the asteroids were initialised on circular orbits. The spread in the curves {\rev in each figure} arise from the different initial values of $a_{\rm ast}$, $i_{\rm ast}$ and the orbital angles. 

\paragraph{\boldmath$R_{\rm ast} = 1000$ km asteroids}

Even in the extreme case of Model A, asteroids or planets with $R_{\rm ast} = 1000$ km (Fig. \ref{Test1410}) experience negligible Yarkovsky drift, and hence represent a useful standard for comparison. The evolution in Fig. \ref{Test1410} is what one would expect without any radiative effects: the inner disc, the planet (black curve), and the outer disc all increase their semimajor axes by a factor of about three based purely on stellar mass loss. The spread seen in the plot primarily mirrors the different semimajor axes chosen in the initial conditions.

\paragraph{\boldmath$R_{\rm ast} = 100$ km asteroids}

For smaller asteroids, with $R_{\rm ast} = 100$~km (Fig. \ref{Test1310}), differences in the evolution become apparent. Both the inner and outer disc drift further outward from the $R_{\rm ast} = 1000$ km case, indicating that the Yarkovsky effect is noticeable. {\rev Sharp changes in slope for inner disc asteroids close to the planet indicate that here planet-asteroid interactions play a more significant role than in Fig. \ref{Test1410}}. At kinks in the inner disc curves, resonant capture occurs, and is maintained. During this resonant capture, the asteroid's eccentricity is increased to a greater extent than can be generated by the Yarkovsky effect. The resonant capture is maintained because these asteroids are too large to ``break free'' from the Yarkovsky effect.

\paragraph{\boldmath$R_{\rm ast} = 10$ km asteroids}

{\rev The evolution of asteroids which are one order of magnitude smaller ($R_{\rm ast} = 10$ km) exhibits planet crossing of the inner disc (Fig. \ref{Test1210}). Even though the planet disrupts the asteroid orbits, the disruption is not great enough to qualitatively alter the final outcome. Nevertheless, the disruption in some cases highlights resonant crossings that were described in Section 2. In the lower panel of Fig. \ref{Test1210}, we have isolated the evolution of five notable cases, which all experience temporary resonant trapping. This trapping occurs when the asteroid semimajor axis suddenly flattens out, or mirrors the planet's evolution, seemingly unaffected by the Yarkovsky effect. These highlighted cases exhibit resonant trapping inward of the planet, outward of the planet, and coincident with the planet.  
}

{\rev The coincident cases are of particular interest because they might showcase entrapment into the planet's Hill sphere. Hence, we investigate further the evolution of the purple curve in Fig. \ref{Test1210} with Fig. \ref{Test1210res}, which highlights the timescale ($\approx 0.38-0.52$ Myr) over which the planet's and asteroid's semimajor axes track each other. The upper panel of Fig. \ref{Test1210res} confirms that the asteroid is temporarily captured into a co-orbital $1$:$1$ mean-motion resonance. The resonant angle $\lambda_{\rm ast}  - \lambda_{\rm pl} + \varpi_{\rm ast} - \varpi_{\rm pl}$ librates with a varying centre and amplitude, where $\lambda$ denotes mean longitude and $\varpi$ denotes longitude of pericentre. Often resonant angles librate around $0^{\circ}$ or $180^{\circ}$, but here the libration centre is here clearly under $180^{\circ}$, perhaps illustrating that a correction term due to the Yarkovsky effect would need to be included in the resonant angle. Computation of this correction term may not be trivial, even for the simplistic case of Model A, because the term could be a function of all orbital elements through Eqs. 59, A5 and A6 of \cite{veretal2015}. During the capture, $e_{\rm ast} \approx 0.15$, but increases steadily, which might trigger the departure from the resonant state.

The bottom panel of Fig. \ref{Test1210res} tracks the mutual distance between the planet and asteroid, which could help indicate if the asteroid becomes trapped within the planet's Hill sphere. Because the Hill sphere for Neptune is under 1 au, the plot shows that the asteroid is not trapped within the Hill sphere, although does occasionally poke into it. The bottom panel's $x$-axis is extended from the $x$-axis of the top panel to show both how the asteroid gradually drifts away from the planet over time and one particularly deep incursion into the Hill sphere at 0.54 Myr, for which there is a corresponding single libration on the top panel.
}

\paragraph{\boldmath$R_{\rm ast} = 1$ km and \boldmath$100$ m asteroids}

For even smaller asteroids in the inner disc (Fig. \ref{Comboa}), the orbit crossing with the planet {\rev occurs too quickly for resonant capture to occur, and the semimajor axes that the asteroids attain along the white dwarf phase is orders of magnitude higher than the planet's}.  Fig. \ref{Comboa} displays the results of {\rev two} simulations on a single plot: those containing asteroids with $R_{\rm ast} = 1$ km and $100$ m. In no case did the planet's interaction with an asteroid trigger instability.  Both sets of inner disc asteroids were propelled to a high enough semimajor axis for eccentricity to start playing a role in nonadiabatic evolution from mass loss \citep{veretal2011}. However, the eventual semimajor axis obtained along the white dwarf phase is primarily due to the Yarkovsky effect. In the $R_{\rm ast} = 100$ m simulation, 81 of the 200 asteroids escaped the system. Here escape refers to leaving the time-dependent Hill ellipsoid of the star, as defined by \cite{vereva2013} and \cite{veretal2014c}, assuming that the star resides in the Solar Neighbourhood (at a distance of 8 kpc from the Galactic Centre).

\paragraph{Effect of planet}

In order to better pinpoint the dependencies of the asteroid evolution on the presence of the planet and the initial eccentricities of the asteroids, we have created a figure (Fig. \ref{Comp1210a}) with four plots that may be compared to one another on the same scale. The upper-left panel reproduces the $R_{\rm ast} = 10$ km curves from Fig. \ref{Test1210} but without the planet curve. {\rev The upper-right plot} shows the same simulation but without a planet. This comparison indicates that the {\rev disruption created by the planet does not qualitatively affect the final result}.  For the case when asteroids have initially eccentric orbits ranging from $e_{\rm ast} = 0.3-0.7$ (bottom panels), then the planet has a marginal effect on the final semimajor axis distribution, despite strongly affecting a few asteroids (like the one associated with the bottom green curve).

\subsubsection{Eccentricity evolution}

\begin{figure*}
\centerline{
\includegraphics[width=8.8cm]{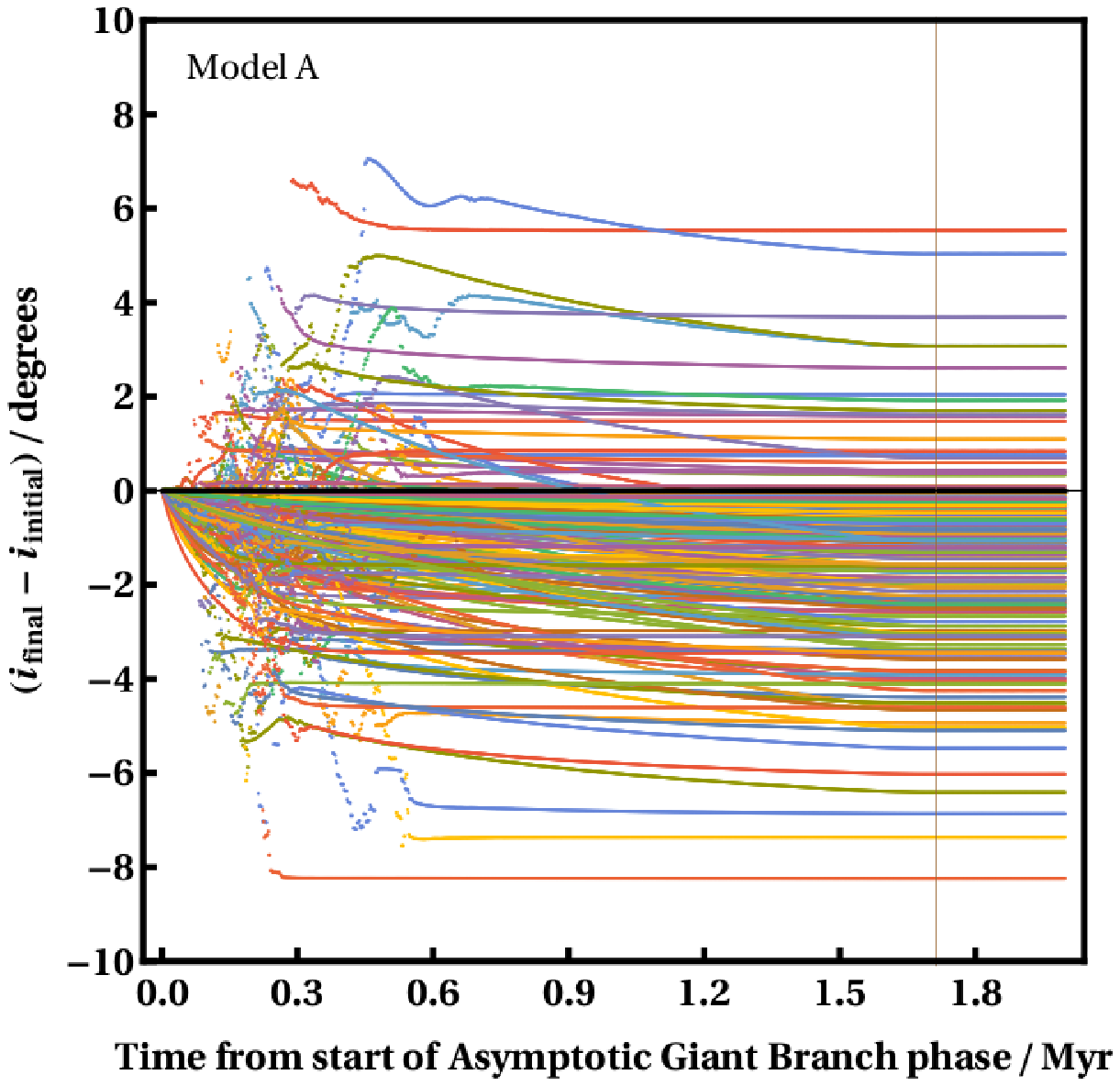}
\ \ \ \ \ \ \ \
\includegraphics[width=8.8cm]{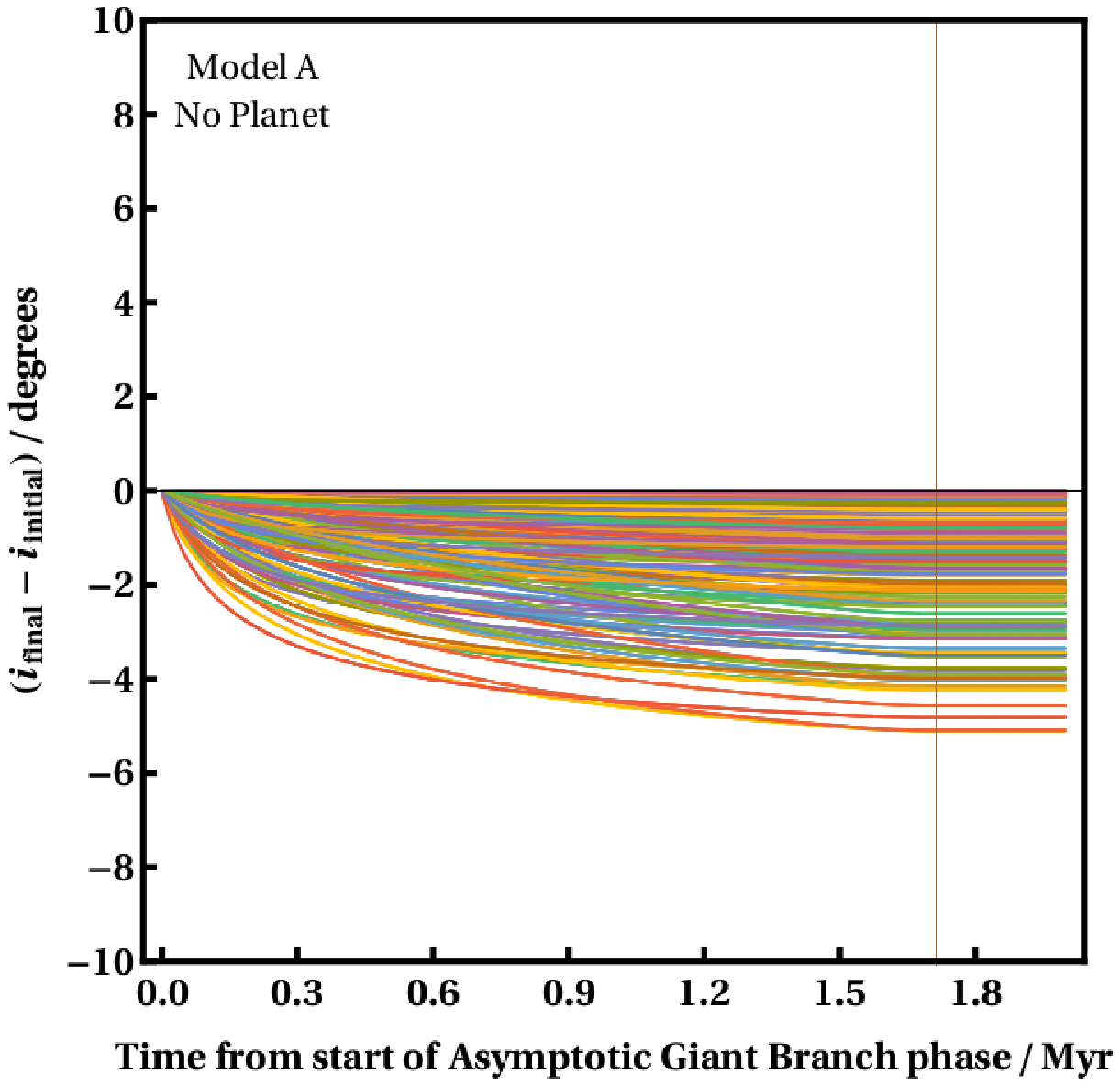}
}
\caption{
The influence of the planet on the inclination evolution of the $R_{\rm ast} = 10$km asteroids
from Fig. \ref{Comp1210e} under {\forref the} Model A Yarkovsky {\forref effect}. 
With no planet (right panel), the
inclination evolves in a direction anticipated by equation (\ref{ievo}). The presence of a planet
(left panel) spreads out the curves. Isotropic stellar mass loss has no effect on inclination evolution.
}
\label{Comp1210i}
\end{figure*}

\begin{figure*}
\centerline{
\includegraphics[width=8.8cm]{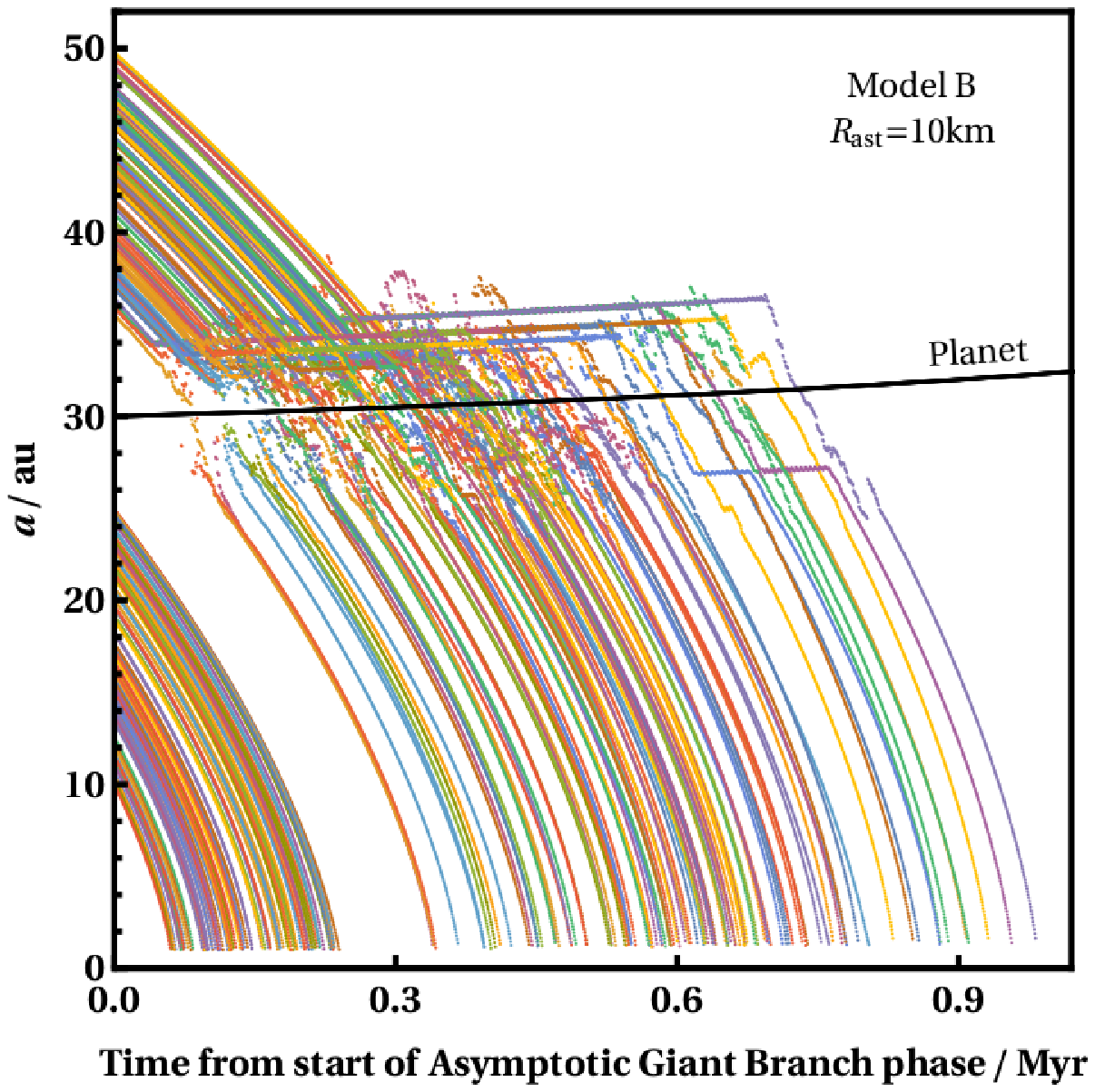}
\ \ \ \ \ \ \ \
\includegraphics[width=8.8cm]{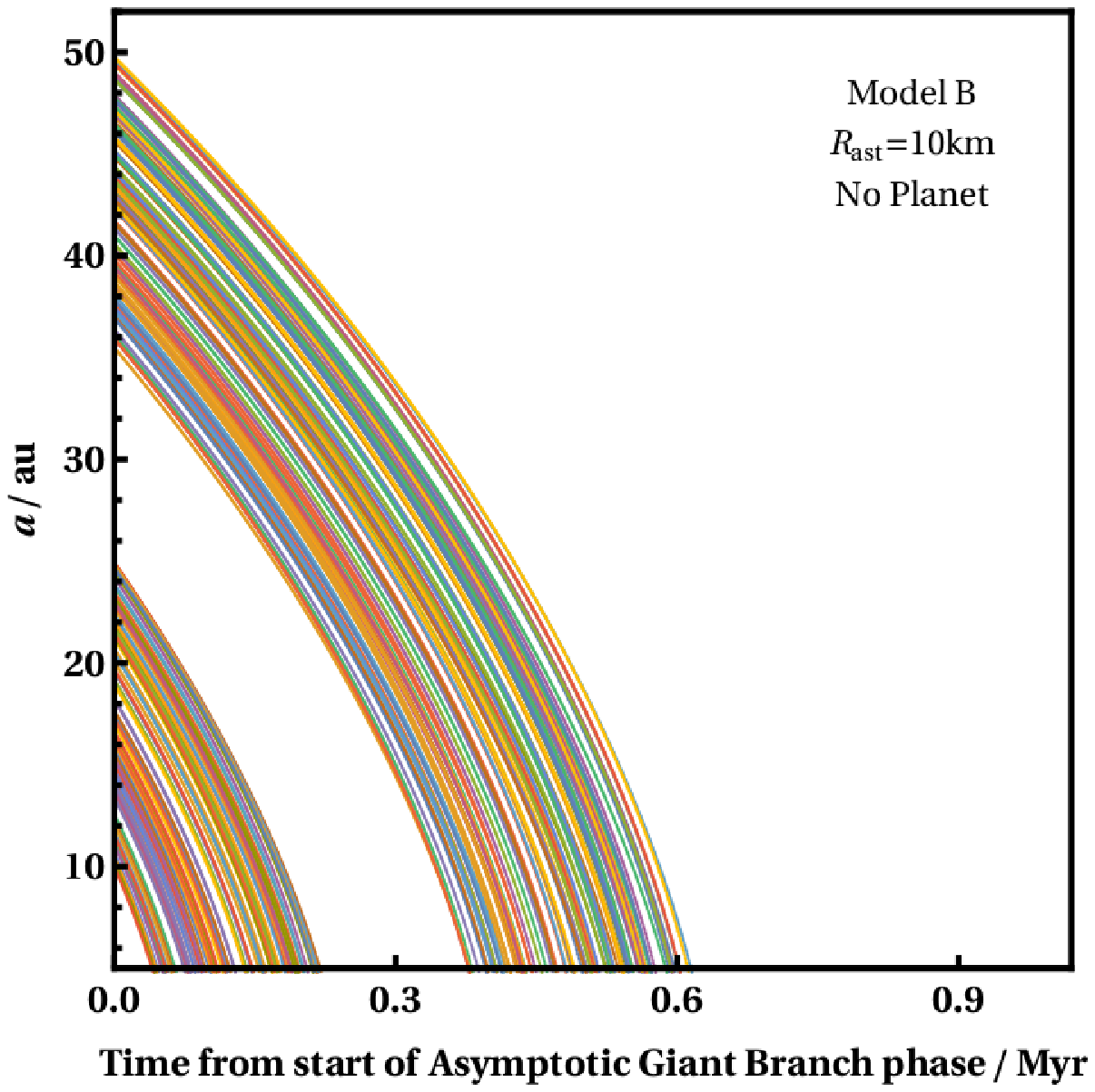}
}
\caption{
How asteroids of size $R_{\rm ast} = 10$ km respond to {\forref the} Model B Yarkovsky {\forref effect}. The left panel includes the presence of a planet, and the right panel does not. {\rev Despite several clear instances of resonant
capture}, all asteroids are eventually engulfed into the star. The presence of the planet does not affect this outcome (despite delaying the engulfment times by a few
$10^5$ yr).
}
\label{Comp1260a}
\end{figure*}

\begin{figure*}
\centerline{
\includegraphics[width=8.8cm]{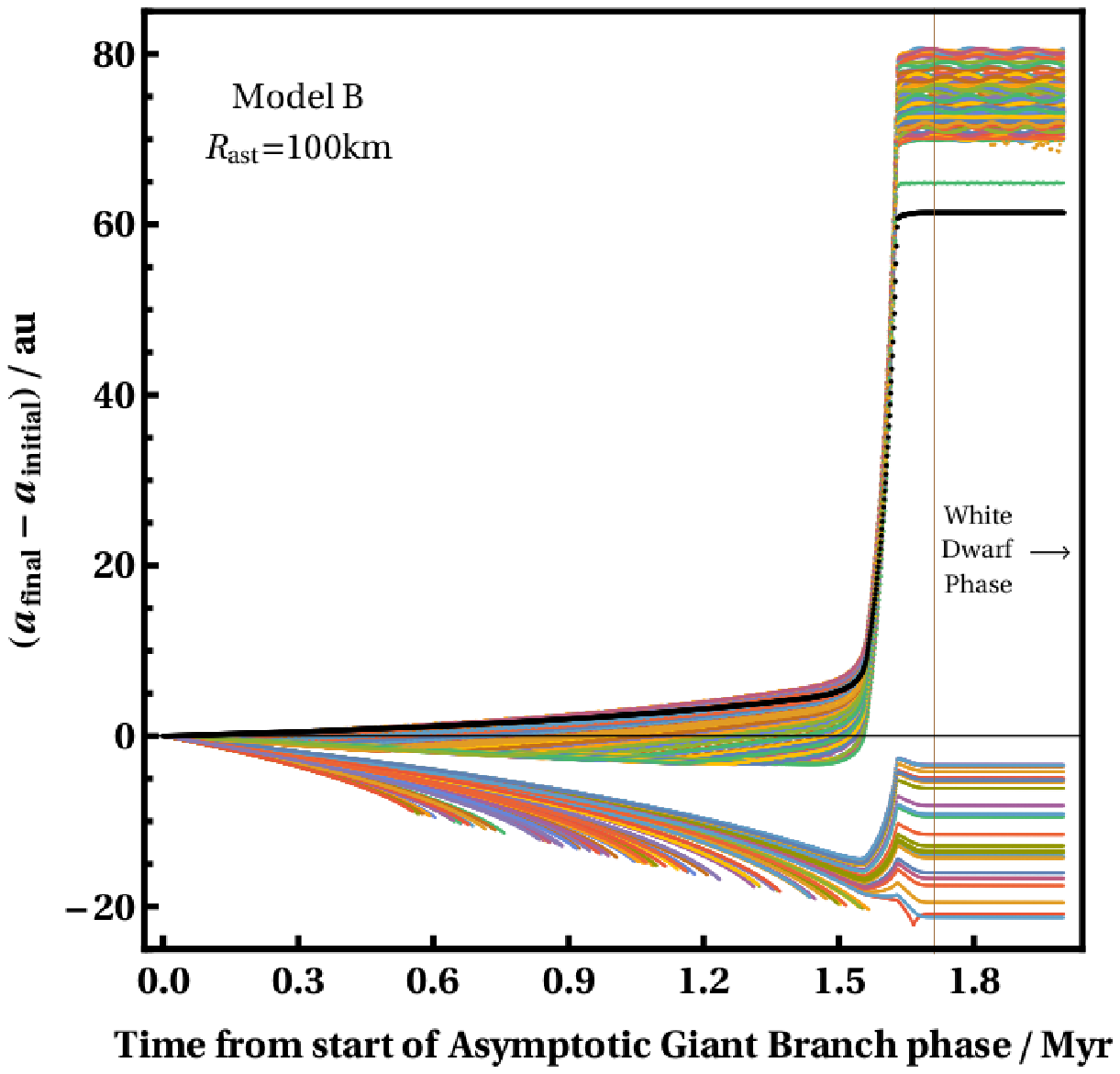}
\ \ \ \ \ \ \ \
\includegraphics[width=8.8cm]{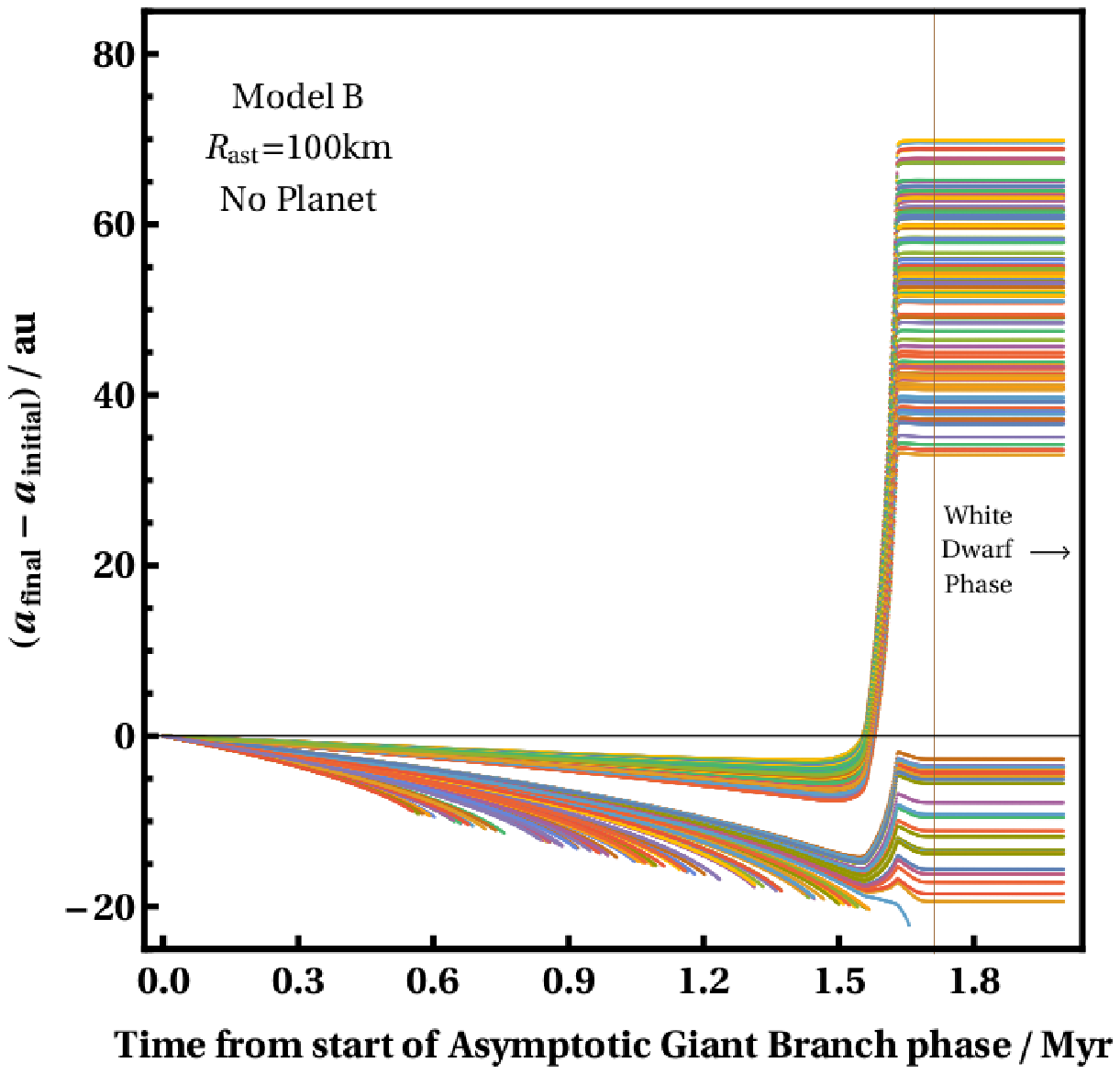}
}
\caption{
Same as Fig. \ref{Comp1260a}, except for $R_{\rm ast} = 100$ km asteroids. In this case, stellar mass
loss and {\forref the} Model B Yarkovsky {\forref effect} compete with each other, leading to net inward drift of the inner disc
and net outward drift of the outer disc. Without the planet, 71/95 inner disc asteroids are engulfed; with
the planet, instead 72/95 are engulfed.
}
\label{Comp1360a}
\end{figure*}

\begin{figure*}
\centerline{
\includegraphics[width=8.8cm]{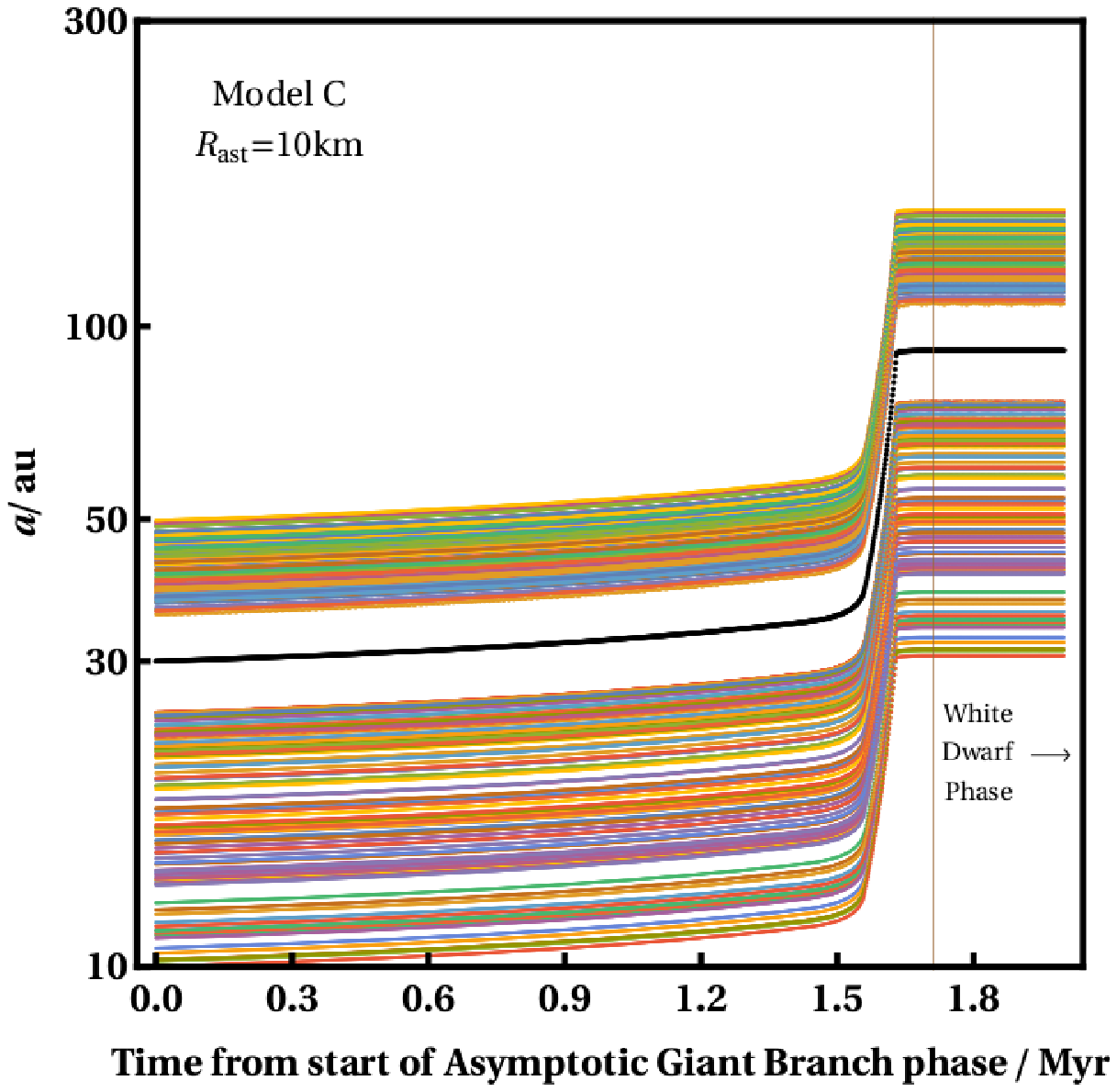}
\ \ \ \ \ \ \ \
\includegraphics[width=8.8cm]{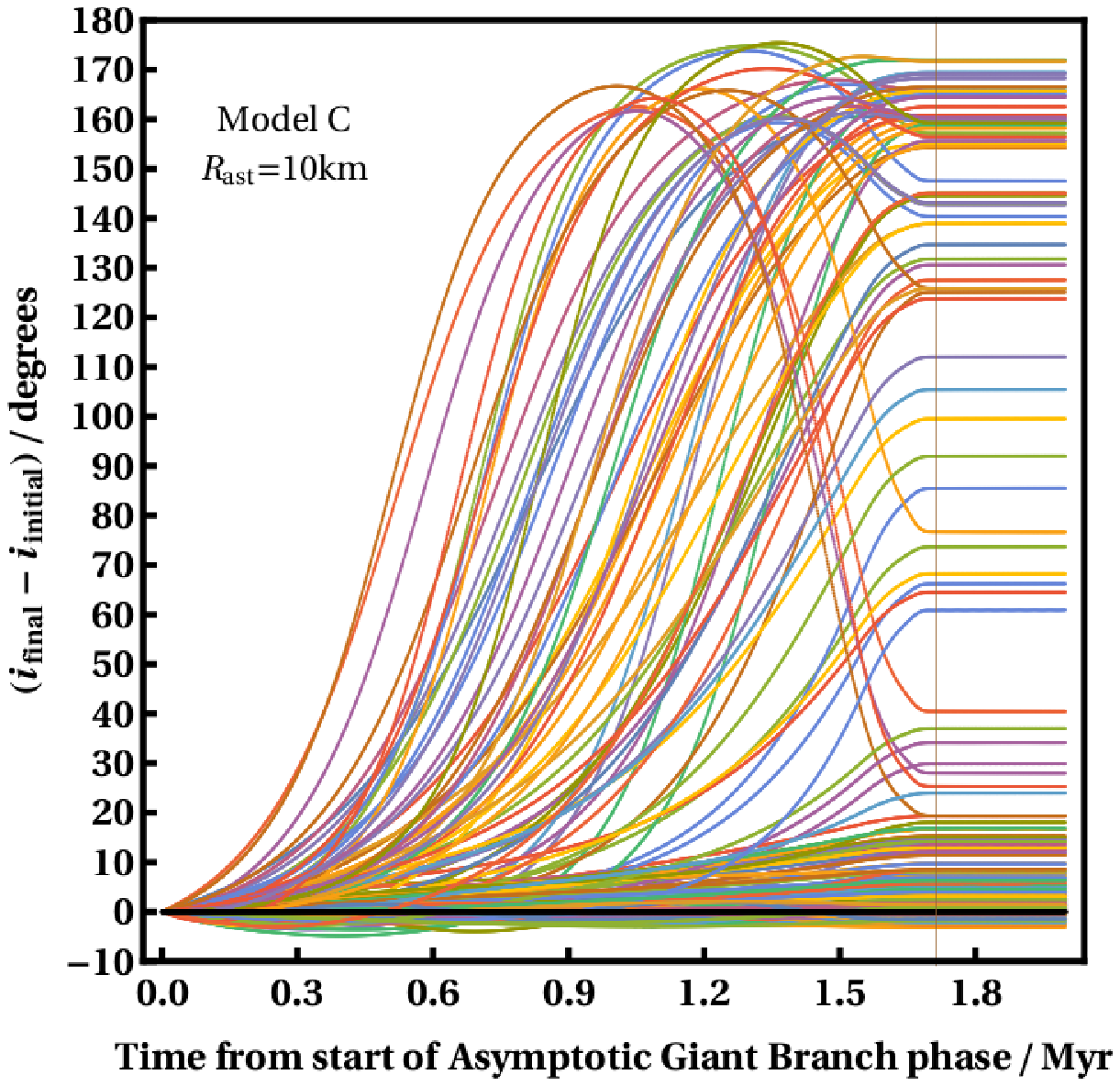}
}
\caption{
Semimajor axis and inclination evolution of asteroids due to {\forref the} Model C Yarkovsky {\forref effect}. This model
completely quenches Yarkovsky-induced semimajor axis drift (left panel), reducing radiative drift to Poynting-Robertson
drag, which is negligible compared to stellar mass loss-induced orbital expansion. However, this Yarkovsky
{\forref effect} can incite inclination evolution (right panel), sometimes easily and smoothly flipping asteroid orbits from prograde
to retrograde and vice versa.
}
\label{Comp1220a}
\end{figure*}

Regarding how the eccentricity of the asteroids themselves change throughout the evolution for Yarkovsky Model A, 
consider Fig. \ref{Comp1210e}, which includes all 200 simulated $R_{\rm ast} = 10$km asteroids from the upper panels
of Fig. \ref{Comp1210a}. The right panel of Fig. \ref{Comp1210e}, which does 
not include a planet, helps confirm equation (\ref{eevo}) by
illustrating that {\forref the radiation-driven} Yarkovsky {\forref effect} does not generate eccentricity in initially circular nearly coplanar asteroids.
Nearly all of the eccentricity shown in this plot is created by stellar mass loss, at the expected level\footnote{The 
division between adiabatic and non-adiabatic evolution from mass loss is not sharp: non-zero eccentricity is always 
generated, even if relatively small, regardless of semimajor axis \citep{veretal2011}.}.

The inclusion of a planet (left panel) dominates the eccentricity evolution of asteroids by orders of magnitude; 
the final eccentricities of most of the asteroids are greater than 0.1. Consequently, a planet exerts greater
influence on the eccentricity evolution of asteroid discs than their semimajor axis evolution.
This eccentricity is pumped almost immediately -- within about $10^5$ yr -- and is independent of post-main-sequence
evolution. Note that the eccentricity of the planet itself (black curve) is also increased due to stellar mass loss
up to about $10^{-2}$.

{\rev
This asteroid eccentricity pumping may be key to understanding why Hill sphere capture does not occur
in our simulations. As indicated by Fig. \ref{CaptEcc}, an increase of just 0.06 may prevent capture from occurring.
Without a protoplanetary disc to keep the asteroid eccentricity low, the gravitational interaction between
the planet and asteroid alone may be strong enough to prevent capture from occurring. Nevertheless,
not all of the asteroid eccentricities are pumped to values higher than 0.06, and the bottom panel of Fig. \ref{Test1210}
indicates that temporary resonant capture may still occasionally occur. Discs are not actually needed 
to induce resonance capture \citep{rayetal2008}, but increase the chances by over an order of magnitude.

Further, the critical eccentricity condition from equation (\ref{crite}) holds only for small eccentricities, and
along with equations (\ref{dadtcritext}) and (\ref{dadtcritint}), does not take into account the perturbations
from the planet. Further, the asteroid's speed is always changing. Capture within the Hill sphere is also more 
likely if there is a quick dissipation mechanism.
In principle, this dissipation may occur from Yarkovsky dissipation as the star transitions into a white dwarf,
but the timing of this event combined with the size of the asteroid would need to be fine-tuned.
}

\subsubsection{Inclination evolution}

In all of our simulations, our asteroids are given initial inclinations with respect to the planet's
orbital plane {\forref between $0^{\circ}$} and $10^{\circ}$. Fig. \ref{Comp1210i} illustrates how the
Model A {\forref Yarkovsky effect} affects the inclination evolution. The 10 km asteroids in the figure
are the same as those in Fig. \ref{Comp1210e} and in the upper panel of Fig. \ref{Comp1210a}.

The right panel of Fig. \ref{Comp1210i} displays a net decrease in all orbital inclinations.
This behaviour is expected from the analytics (equation \ref{ievo}). Note additionally that
mass loss has no effect on inclination evolution unless the mass loss is anisotropic
\citep{veretal2011,veretal2013a,doskal2016a,doskal2016b,veretal2018}. Therefore, the 
levelling-out of the inclination
evolution curves entirely reflects the reduced influence of the Model A Yarkovsky {\forref effect}
as the asteroid semimajor axes increase in a nonlinear fashion.

The presence of a planet (left panel) spreads the curves about the $0^{\circ}$ line, although
there still remains a net decrease in the inclination evolution. Like for orbital eccentricity, 
the initial perturbations, as seen
up to about 0.6 Myr after the start of the asymptotic giant branch phase, would have realistically occurred
much earlier in the system evolution and is just an artefact of our chosen starting time.

\subsection{Model B}

In Model B, $\mathbb{Q}$ has been chosen to maximise the asteroid's inward drift due the Yarkovsky effect. Because of the dependence of the Yarkovsky drift on $a_{\rm ast}$, there is a positive feedback effect, accelerating the asteroid more quickly towards the central star as the semimajor axis decreases. However, stellar mass loss represents a competing effect, pushing the orbit outward. 

Which effect wins depends on the size of the asteroid.  Outward expansion due to mass loss is independent of the asteroid mass at the $M_{\rm ast}/M_{\star}$ level, whereas {\forref the} Yarkovsky {\forref effect} is proportional to the square of the asteroid radius (equation \ref{yarfi2}). This latter strong dependence dictates that asteroids with radii as large as $10$ km cannot avoid engulfment, regardless of the presence of the planet.  

Figure \ref{Comp1260a} illustrates their evolution, for asteroids on initially circular orbits, and showcases the potentially destructive power of {\forref the radiation-induced Yarkovsky effect}. In each case, the evolution of all 200 asteroids are shown, and none survive to reach the tip of the Asymptotic Giant Branch phase. The left panel {\rev includes the influence of the planet, and the right panel does not}. Although the planet disperses incoming asteroid orbits with a flair apparent in the left panel, the presence of the planet does not actually qualitatively change the fate of these asteroids. Near-horizontal segments in the left panel indicate temporary capture within mean motion resonance, which can last long enough to delay engulfment into the star for a few $10^5$ yr. We do not model engulfment nor tidal effects, as none of these asteroid evolutions are necessarily realistic anyway: recall they represent just bounding cases for Yarkovsky {\forref effects} along the giant branch evolution phases.

For $R_{\rm ast} = 100$ km asteroids, the competing effects of Model B Yarkovsky {\forref drift} and stellar mass loss are comparable (Fig. \ref{Comp1360a}). {\rev Shown is the semimajor axis change from its initial value, to highlight decreases versus increases.} In fact, the direction of semimajor axis drift of one of these asteroids changes depending on if it resides in the inner or outer disc. Figure \ref{Comp1360a} illustrates the evolution of both discs, with (left panel) and without (right panel) a planet. The inner disc of 95 asteroids is dragged towards the star, where 71 become engulfed. All 105 asteroids in the outer disc survive, despite initially being slightly dragged towards the star before increasing their semimajor axes. The presence of the planet slightly helps prop up the semimajor axes of the outer disc, but again does not qualitatively affect the outcome. In fact, the planet induces just one more asteroid to be engulfed than if the planet was not present.

\subsection{Model C}

In Model C, we quench the planar component
of all Yarkovsky-based semimajor axis evolution by setting $\mathbb{Q}_{12} =  \mathbb{Q}_{21}$. 
The model also zeroes out the other non-diagonal terms
of $\mathbb{Q}$
($\mathbb{Q}_{13}=\mathbb{Q}_{23}=\mathbb{Q}_{31}=\mathbb{Q}_{32}$) which 
reverts the semimajor axis drift to Poynting-Robertson drag.

Confirmation of this quenching appears in the left panel of Fig. \ref{Comp1220a}, which shows the discs evolving 
along the giant branch due primarily to mass loss. We have also tested semimajor axis evolution for both $R_{\rm ast} = 1$ km
and $R_{\rm ast} = 100$m asteroids in low-resolution simulations with just a few asteroids each (for computational feasibility), 
and the result is the same. 

The right panel of the figure, however,
reveals that the inclination evolution is not zeroed-out, and in fact can change significantly more 
than in Model A. The reason is because for circular, near-coplanar orbits in Model C,

\begin{equation}
\left \langle \frac{di_{\rm ast}}{dt} \right \rangle \propto \frac{
\sin{i} \cos{\left(2\Omega_{\rm ast}\right)}  
}{a_{\rm ast}^{3/2}}
,
\label{ievo2}
\end{equation}

\noindent{}which allows the inclination to change initially more quickly than in
equation (\ref{ievo}). Another reason is that $\left \langle \frac{d\Omega_{\rm ast}}{dt} \right \rangle$ 
is exactly twice as fast as in Model A in this limit. {\rev When both of these differential equations are
solved simultaneously, time can be eliminated, yielding}

\begin{equation}
i_{\rm ast}|_{\rm Model \ A} \approx \sin^{-1} \left[ \frac{C}{\sin{\Omega_{\rm ast}}} \right],
\label{Ainc}
\end{equation}

\begin{equation}
i_{\rm ast}|_{\rm Model \ C} \approx \sin^{-1} \left[ \frac{C}{\sqrt{\sin{\left(2\Omega_{\rm ast}\right)}}} \right],
\label{Cinc}
\end{equation}

{\rev
\noindent{}where $C$ is an arbitrary constant. Equations (\ref{Ainc}-\ref{Cinc})
explicitly illustrate the steeper dependence on longitude of ascending node in Model C}. 

These initial quick changes can have a positive feedback effect on 
the asteroid inclination before $a_{\rm ast}$ becomes too large from stellar mass loss.
{\rev Note that the right panel of Fig. \ref{Comp1220a} shows that only some asteroids
achieve inclinations of several tens of degrees, because of the spread of their initial values. Further,
all of the asteroids which ended up on the most highly inclined orbits had a ``quick start'' (fast initial
inclination change). In this
sense, the duration of the giant branch phase is key to this inclination pumping. Here we sampled just one case, with an
initially $2M_{\odot}$ star. More massive stars would have shorter giant branch lifetimes but
higher luminosities.} 

Model C Yarkovsky 
{\forref drift} illustrates how an asteroid's orbital inclination can flip from prograde to retrograde
and back again in a smooth fashion. {\rev One potential consequence of this type of motion 
is the formation of a near-isotropic cloud of debris, perhaps akin to so-called mini Oort-clouds
\citep{rayarm2013}. The geometry of these clouds could have potentially important implications for
white dwarf pollution \citep{alcetal1986,paralc1998,veretal2014d,stoetal2015,caihey2017}}.

\section{Discussion}

None of the individual asteroid evolutions that we simulated are likely to be realistic. 
{\rev The reality is complex, as indicated both by observations from within the solar system
and by detailed efforts to model those observations.} Rather, our simulations 
reveal physical interplays and provide limits on the characteristics of motion that we might expect for given parameters.
These approximations do, however, give some more depth to the constant semimajor axis drift 
approximations sometimes employed
for asteroids orbiting the Sun; this generalisation is necessary in giant branch planetary systems
because of the host star's ``supercharged'' luminosity.

However, our models do not go nearly far enough. The entries of $\mathbb{Q}$ are functions of time, and
in turn, of the spin and thermal properties of the asteroid \citep{veretal2015}. The asteroid's spin and thermal properties may
then be a strong function of its shape \citep{rozgre2012,rozgre2013,voketal2015,goletal2016}, which, as indicated by the majority of solar system asteroids, is unlikely
to be spherical. 

Another consideration is the connection between the Yarkovsky and YORP effects. {\forref YORP spin-down can lead
to tumbling states (the asteroid Toutatis is an example of a tumbler), and these states can ``switch off'' the Yarkovsky effect
\citep{voketal2007}. However, tumbling states could represent transient phenomena in systems where host stars change their
luminosity on short timescales. Without dedicated investigations, this interplay remains unclear.  Further,} giant branch rotational
fission from YORP could be ubiquitous within a particular distance (estimated to be about 7 au by \citealt*{veretal2014a}),
and the resulting fragments will be closer to the giant branch star than any considered here, and hence subject
to even stronger Yarkovsky forces. Further, some asteroids at Kuiper-belt like distances, or further away, may already be spinning
quickly enough to experience YORP break-up during giant branch evolution.

{\forref Our model ignores the effect of collisions, which can play a vital role in shaping the debris fields of giant branch planetary systems \citep{bonwya2010}. These collisions can both modify the size distribution of the asteroids towards smaller sizes, and affect YORP evolution \citep{maretal2011} through the resetting of the spin state (with a timescale that goes as the square root of the asteroid size, see e.g. \citealt*{faretal1998}). A driver for collisional activity is eccentricity excitation, which we have shown is generated by the presence of a planet (Fig. \ref{Comp1210e}). The consequence of significant collisional comminution will be an increase in the type of behaviour seen in Fig. \ref{Comboa}, where smaller particles are flung into the outer reaches of the planetary system. The final orbital distribution of debris may then tend towards high values of semimajor axis when collisions are frequent and early. In this respect, by not modelling collisions, our simulations provide a type of lower bound for the extent of the eventual outward migration of debris.}

A helpful result of this paper is that {\forref the} giant branch Yarkovsky {\forref effect} can be strong enough to render
the presence of a planet as relatively unimportant. Planets certainly play a dynamical role in pumping eccentricity
(Fig. \ref{Comp1210e}) and changing timescales for destruction (Fig. \ref{Comp1260a}). However, they 
struggle to retain asteroids in mean motion resonances and, in the big picture, do not qualitatively
change how many asteroids are destroyed around giant branch stars or where asteroids will reside 
after reaching the white dwarf phase. 

These statements, however, must be caveated with the fact that 
dissipation at different points during the evolution could result in temporary capture 
into mean motion resonances (Sections 2.1 {\rev and 4.1.4}) or within the Hill sphere of the planet
(\citealt*{higida2016,higida2017}; also see Sections 2.2-2.3).

{\rev The relevant formulae (equations  \ref{dadtcritext}, \ref{dadtcritint}, \ref{crite}, \ref{atc} and \ref{etc}) 
are affected by post-main-sequence evolution through just $a_{\rm pl}$, $a_{\rm ast}$ and $M_{\star}$. The changes in these critical values that are induced
by variations in $M_{\star}$ are no greater than a factor of a few. This factor of a few change also holds for $a_{\rm pl}$ and $a_{\rm ast}$ unless
the latter value becomes so large that the planet would be too far away to produce capture anyway. Planets could play a much larger
role with a time-varying $\mathbb{Q}$, which may provide dissipation at key times.
}

Another useful result of this paper are constraints on asteroid sizes. Even in the most extreme cases 
(exemplified by Models A, B, and C), the largest exo-Kuiper belt asteroids -- on the order of 
$R_{\rm ast} = 1000$ km -- are not meaningfully affected by {\forref the} Yarkovsky {\forref effect}. On the other end
of the size spectrum, almost every asteroid smaller than about 10 km is predominately perturbed by
{\forref the radiation-induced Yarkovsky effect}. This effect should be considered when establishing initial conditions for simulations along the white dwarf phase.

As is apparent in most of the plots in this paper, dynamical settling occurs quickly just at the onset
of the white dwarf phase. Although the Yarkovsky {\forref effect} does not ``turn off'' due to the white dwarf,
the luminosity of the white dwarf becomes {\forref sub-solar} within just a few Myr. Combined with the greatly
expanded asteroid orbits from giant branch evolution, the white {\forref dwarf radiation-induced} 
Yarkovsky {\forref effect} is then often negligible.

In this paper, we sampled only a brief time interval within the entire evolution of a planetary system (due to computational limitations). Hence, our initial conditions were not realistic, but were chosen instead to demonstrate physical trends. Reaching the asymptotic giant branch phase would have required asteroids to first undergo all of main sequence and giant branch evolution, and maintaining  $e_{\rm ast} = 0$ throughout those periods may be difficult. During the white dwarf phase, instabilities amongst planets coupled with planet-asteroid interactions can lead to white dwarf pollution \citep{bonetal2011,debetal2012,frehan2014,veretal2016,musetal2018,smaetal2018}. Therefore, accurately computing the relative positions of the asteroids and planets after the asymptotic giant branch phase is crucial for the estimation of white dwarf pollution rates and timescales.

Will our own Sun be polluted? The answer largely depends on the fate of the Kuiper Belt, along with whether Planet Nine exists \citep{veras2016b}. One complication for the solar system that is not addressed in this paper is the Yarkovsky {\forref effect} generated by a red giant branch star (as opposed to an asymptotic giant branch star). $1M_{\odot}$ main sequence stars will undergo two largely comparable periods of enhanced luminosity, at the tips of both the red giant and asymptotic giant branch phases. The flavour of Yarkovsky {\forref drift} (Model A versus Model B versus Model C versus something in-between) could change both within and between these two phases.

\section{Summary}

Understanding how asteroids evolve during giant branch evolution crucially determines their capability to pollute the eventual white dwarf. Although the often dominant effects of Yarkvosky {\forref drift} generated from post-main-sequence {\forref stellar radiation} have been analytically estimated previously \citep{veretal2015}, here we provide more detail by (i) running numerical simulations with three different {\rev extreme} models, and (ii) introducing a planet. The three models were chosen to place limits on the types of motion and orbital changes that we may expect, and could aid the future development of more sophisticated shape, spin and thermal inertia constructions. The range of outcomes is much greater than what is observed in the solar system, with semimajor axis changes that can vary by orders of magnitude and easily achieved orbital inclination flipping. Amidst this enhanced radiative forcing, the influence of the planet is minimised. We also analytically considered how these planets could capture asteroids within their Hill spheres \citep{higida2016,higida2017} and into mean motion resonances (as in protoplanetary disc migration), but numerically found that these processes are {\rev infrequent} and ineffectual without a {\rev fine-tuned} dissipation prescription.

\section*{Acknowledgements}

{\forref We are grateful to Apostolos Christou for his helpful referee report, which has improved the manuscript.} DV thanks the Earth-Life Science Institute at the Tokyo Institute of Technology for their hospitality during his stay and for the initiation of this project. DV also gratefully acknowledges the support of the STFC via an Ernest Rutherford Fellowship (grant ST/P003850/1).



\label{lastpage}

\begin{thebibliography}{99}


\bibitem[Adams et al.(2008)]{adalau2008} Adams, F.~C., Laughlin, G. \& Bloch, A.~M.\ 2008, ApJ, 683, 1117.

\bibitem[Adams \& Bloch(2013)]{adablo2013} Adams, F.~C., \& Bloch, A.~M.\ 2013, ApJL, 777, L30 

\bibitem[Adams et al.(2013)]{adaetal2013} Adams, F.~C., Anderson, K.~R., \& Bloch, A.~M.\ 2013, MNRAS, 432, 438 

\bibitem[Alcock et al.(1986)]{alcetal1986} Alcock, C., Fristrom, C.C., \& Siegelman, R.\ 1986, ApJ, 302, 462 






\bibitem[Batygin(2015)]{batygin2015} Batygin, K.\ 2015, MNRAS, 451, 2589



\bibitem[Bodman \& Quillen(2014)]{bodqui2014} Bodman, E.~H.~L. \& Quillen, A.~C.\ 2014, MNRAS, 440, 1753.

\bibitem[Bonsor \& Wyatt(2010)]{bonwya2010} Bonsor, A., \& Wyatt, M.\ 2010, MNRAS, 409, 1631 

\bibitem[Bonsor et al.(2011)]{bonetal2011} Bonsor, A., Mustill, A.~J., \& Wyatt, M.~C.\ 2011, MNRAS, 414, 930 



\bibitem[Bottke et al.(2000)]{botetal2000} Bottke, W.~F., Jr., Rubincam, D.~P., \& Burns, J.~A.\ 2000, Icarus, 145, 301 

\bibitem[Bottke et al.(2006)]{botetal2006} Bottke, W.~F., Jr., Vokrouhlick{\'y}, D., Rubincam, D.~P., \& Nesvorn{\'y}, D.\ 2006, Annual Review of Earth and Planetary Sciences, 34, 157 
  


\bibitem[Bro{\v z}(2006)]{broz2006} Bro{\v z}, M.\ 2006, Ph.D.~Thesis,  
  

\bibitem[Caiazzo \& Heyl(2017)]{caihey2017} Caiazzo, I., \& Heyl, J.~S.\ 2017, MNRAS, 469, 2750


\bibitem[Chambers(1999)]{chambers1999} Chambers, J.~E.\ 1999, MNRAS, 304, 793 

\bibitem[Cresswell \& Nelson(2008)]{crenel2008} Cresswell, P., \& Nelson, R.~P.\ 2008, A\&A, 482, 677 



\bibitem[Debes \& Sigurdsson(2002)]{debsig2002} Debes, J.~H., \& Sigurdsson, S.\ 2002, ApJ, 572, 556 

\bibitem[Debes et al.(2012)]{debetal2012} Debes, J.~H., Walsh, K.~J., \& Stark, C.\ 2012, ApJ, 747, 148 

\bibitem[Deck \& Batygin(2015)]{decbat2015} Deck, K.~M., \& Batygin, K.\ 2015, ApJ, 810, 119


\bibitem[Dong et al.(2010)]{donetal2010} Dong, R., Wang, Y., Lin, D.~N.~C., \& Liu, X.-W.\ 2010, ApJ, 715, 1036 

\bibitem[Dosopoulou \& Kalogera(2016a)]{doskal2016a} Dosopoulou, F., \& Kalogera, V.\ 2016a, ApJ, 825, 70 

\bibitem[Dosopoulou \& Kalogera(2016b)]{doskal2016b} Dosopoulou, F., \& Kalogera, V.\ 2016b, ApJ, 825, 71 




\bibitem[El Moutamid, et al.(2017)]{elmetal2017} El Moutamid, M., Sicardy, B. \& Renner, S.\ 2017, MNRAS, 469, 2380.







\bibitem[Farihi(2016)]{farihi2016} Farihi, J.\ 2016, New Astronomy Reviews, 71, 9 


\bibitem[Farinella et al.(1998)]{faretal1998} Farinella, P., Vokrouhlick{\'y}, D., \& Hartmann, W.~K.\ 1998, Icarus, 132, 378 

\bibitem[Folonier et al.(2014)]{foletal2014} Folonier, H.~A., Roig, F., \& Beaug{\'e}, C.\ 2014, Celestial Mechanics and Dynamical Astronomy, 119, 1  

  

\bibitem[Frewen \& Hansen(2014)]{frehan2014} Frewen, S.~F.~N., \& Hansen, B.~M.~S.\ 2014, MNRAS, 439, 2442 

\bibitem[Friedland(2001)]{friedland2001} Friedland, L.\ 2001, ApJL, 547, L75 


\bibitem[Gallardo et al.(2011)]{galetal2011} Gallardo, T., Venturini, J., Roig, F., \& Gil-Hutton, R.\ 2011, Icarus, 214, 632

\bibitem[Gallet et al.(2017)]{galetal2017} Gallet, F., Bolmont, E., Mathis, S., Charbonnel, C., \& Amard, L.\ 2017, In Press A\&A, arXiv:1705.10164 



\bibitem[G{\"a}nsicke et al.(2012)]{gaeetal2012} G{\"a}nsicke, B.~T., Koester, D., Farihi, J., et al.\ 2012, MNRAS, 424, 333 





\bibitem[Goldreich \& Schlichting(2014)]{golsch2014} Goldreich, P. \& Schlichting, H.~E.\ 2014, AJ, 147, 32.

\bibitem[Golubov et al.(2016)]{goletal2016} Golubov, O., Kravets, Y., Krugly, Y.~N., \& Scheeres, D.~J.\ 2016, MNRAS, 458, 3977


\bibitem[Hadjidemetriou(1963)]{hadjidemetriou1963} Hadjidemetriou, J.~D.\ 1963, Icarus, 2, 440 




\bibitem[Hands \& Alexander(2018)]{hanale2018} Hands, T.~O. \& Alexander, R.~D.\ 2018, MNRAS, 474, 3998.

\bibitem[Harrison et al.(2018)]{haretal2018} Harrison, J., submitted to MNRAS



\bibitem[Higuchi \& Ida(2016)]{higida2016} Higuchi, A., \& Ida, S.\ 2016, AJ, 151, 16 
  
\bibitem[Higuchi \& Ida(2017)]{higida2017} Higuchi, A., \& Ida, S.\ 2017, AJ, 153, 155 



\bibitem[Hollands et al.(2018)]{holetal2018} Hollands, M.~A., G{\"a}nsicke, B.~T., \& Koester, D.\ 2018, MNRAS, 477, 93 

\bibitem[Hurley et al.(2000)]{huretal2000} Hurley, J.~R., Pols, O.~R., \& Tout, C.~A.\ 2000, MNRAS, 315, 543 



\bibitem[Jura(2003)]{jura2003} Jura, M.\ 2003, ApJL, 584, L91 



\bibitem[Jura(2008)]{jura2008} Jura, M.\ 2008, AJ, 135, 1785 
  


\bibitem[Jura et al.(2012)]{juretal2012} Jura, M., Xu, S., Klein, B., Koester, D., \& Zuckerman, B.\ 2012, ApJ, 750, 69 

\bibitem[Jura \& Young(2014)]{juryou2014} Jura, M., \& Young, E.~D.\ 2014, Annual Review of Earth and Planetary Sciences, 42, 45 






\bibitem[Klein et al.(2010)]{kleetal2010} Klein, B., Jura, M., Koester, D., Zuckerman, B., \& Melis, C.\ 2010, ApJ, 709, 950 

\bibitem[Klein et al.(2011)]{kleetal2011} Klein, B., Jura, M., Koester, D., \& Zuckerman, B.\ 2011, ApJ, 741, 64 


  

\bibitem[Koester et al.(2014)]{koeetal2014} Koester, D., G{\"a}nsicke, B.~T., \& Farihi, J.\ 2014, A\&A, 566, A34 


\bibitem[Kunitomo et al.(2011)]{kunetal2011} Kunitomo, M., Ikoma, M., Sato, B., Katsuta, Y., \& Ida, S.\ 2011, ApJ, 737, 66 

\bibitem[Lee \& Peale(2002)]{leepea2002} Lee, M.~H., \& Peale, S.~J.\ 2002, ApJ, 567, 596 

\bibitem[Li et al.(2014)]{lietal2014} Li, G., Naoz, S., Valsecchi, F., Johnson, J.~A., \& Rasio, F.~A.\ 2014, ApJ, 794, 131 

\bibitem[Livio \& Soker(1984)]{livsok1984} Livio, M., \& Soker, N.\ 1984, MNRAS, 208, 763 

\bibitem[Luan(2014)]{luan2014} Luan, J.\ 2014, arXiv:1410.2648

\bibitem[Madappatt et al.(2016)]{madetal2016} Madappatt, N., De Marco, O., \& Villaver, E.\ 2016, MNRAS, 463, 1040 

\bibitem[Malamud \& Perets(2016)]{malper2016} Malamud, U., \& Perets, H.~B.\ 2016, ApJ, 832, 160 

\bibitem[Malamud \& Perets(2017a)]{malper2017a} Malamud, U., \& Perets, H.~B.\ 2017a, ApJ, 842, 67 

\bibitem[Malamud \& Perets(2017b)]{malper2017b} Malamud, U., \& Perets, H.~B.\ 2017b, ApJ, 849, 8 

\bibitem[Manser et al.(2016)]{manetal2016} Manser, C.~J., G{\"a}nsicke, B.~T., Marsh, T.~R., et al.\ 2016, MNRAS, 455, 4467 

\bibitem[Manser et al.(2018)]{manetal2018} Manser, C.~J. et al.\ 2018, Submitted

\bibitem[Marzari et al.(2011)]{maretal2011} Marzari, F., Rossi, A., \& Scheeres, D.~J.\ 2011, Icarus, 214, 622



\bibitem[Murray \& Dermott(1999)]{murder1999} Murray, C.~D., \& Dermott, S.~F.\ 1999, Solar system dynamics, Cambridge, UK: Cambridge University Press  
  
\bibitem[Murray-Clay \& Chiang(2006)]{murchi2006} Murray-Clay, R.~A. \& Chiang, E.~I.\ 2006, ApJ, 651, 1194.

\bibitem[Mustill \& Wyatt(2011)]{muswya2011} Mustill, A.~J., \& Wyatt, M.~C.\ 2011, MNRAS, 413, 554 

\bibitem[Mustill \& Wyatt(2012)]{muswya2012} Mustill, A.~J. \& Wyatt, M.~C.\ 2012, MNRAS, 419, 3074.
  
\bibitem[Mustill \& Villaver(2012)]{musvil2012} Mustill, A.~J., \& Villaver, E.\ 2012, ApJ, 761, 121 

\bibitem[Mustill et al.(2014)]{musetal2014} Mustill, A.~J., Veras, D., \& Villaver, E.\ 2014, MNRAS, 437, 1404 

\bibitem[Mustill et al.(2018)]{musetal2018} Mustill, A.~J., Villaver, E., Veras, D., G{\"a}nsicke, B.~T., 
\& Bonsor, A.\ 2018, MNRAS, 476, 3939 

\bibitem[Nelemans \& Tauris(1998)]{neltau1998} Nelemans, G., \& Tauris, T.~M.\ 1998, A\&A, 335, L85 

\bibitem[Nordhaus \& Spiegel(2013)]{norspi2013} Nordhaus, J., \& Spiegel, D.~S.\ 2013, MNRAS, 432, 500 

\bibitem[Ogihara \& Ida(2012)]{ogiida2012} Ogihara, M. \& Ida, S.\ 2012, ApJ, 753, 60.

\bibitem[Ogihara \& Kobayashi(2013)]{ogikob2013} Ogihara, M. \& Kobayashi, H.\ 2013, ApJ, 775, 34.

\bibitem[Omarov(1962)]{omarov1962} Omarov, T.~B. 1962, Izv. Astrofiz. Inst. Acad. Nauk. KazSSR, 14, 66

\bibitem[Pan \& Schlichting(2017)]{pansch2017} Pan, M., \& Schlichting, H.~E.\ 2017, arXiv:1704.07836

\bibitem[Papaloizou \& Szuszkiewicz(2005)]{papszu2005} Papaloizou, J.~C.~B. \& Szuszkiewicz, E.\ 2005, MNRAS, 363, 153.

\bibitem[Parriott \& Alcock(1998)]{paralc1998} Parriott, J., \& Alcock, C.\ 1998, ApJ, 501, 357 




\bibitem[Petrovich et al.(2013)]{petetal2013} Petrovich, C., Malhotra, R., \& Tremaine, S.\ 2013, ApJ, 770, 24 



\bibitem[Polishook et al.(2017)]{poletal2017} Polishook, D., Moskovitz, N., Thirouin, A., et al.\ 2017, Icarus, 297, 126 

\bibitem[Quillen(2006)]{quillen2006} Quillen, A.~C.\ 2006, MNRAS, 365, 1367.

\bibitem[Quillen \& Faber(2006)]{quifab2006} Quillen, A.~C., \& Faber, P.\ 2006, MNRAS, 373, 124

\bibitem[Quillen, et al.(2013)]{quietal2013} Quillen, A.~C., Bodman, E. \& Moore, A.\ 2013, MNRAS, 435, 2256.

\bibitem[Quillen \& French(2014)]{quifre2014} Quillen, A.~C. \& French, R.~S.\ 2014, MNRAS, 445, 3959.

\bibitem[Raymond et al.(2008)]{rayetal2008} Raymond, S.~N., Barnes, R., Armitage, P.~J., \& Gorelick, N.\ 2008, ApJL, 687, L107 

\bibitem[Raymond \& Armitage(2013)]{rayarm2013} Raymond, S.~N., \& Armitage, P.~J.\ 2013, MNRAS, 429, L99










\bibitem[Rein \& Papaloizou(2009)]{reipap2009} Rein, H. \& Papaloizou, J.~C.~B.\ 2009, A\&A, 497, 595.

\bibitem[Rein \& Papaloizou(2010)]{reipap2010} Rein, H. \& Papaloizou, J.~C.~B.\ 2010, A\&A, 524, A22.
  
\bibitem[Rozitis \& Green(2012)]{rozgre2012} Rozitis, B., \& Green, S.~F.\ 2012, MNRAS, 423, 367 

\bibitem[Rozitis \& Green(2013)]{rozgre2013} Rozitis, B., \& Green, S.~F.\ 2013, MNRAS, 433, 603 



\bibitem[Schr{\"o}der \& Connon Smith(2008)]{schcon2008} Schr{\"o}der, K.-P., \& Connon Smith, R.\ 2008, MNRAS, 386, 155 

\bibitem[Shannon, et al.(2015)]{shaetal2015} Shannon, A., Mustill, A.~J. \& Wyatt, M.\ 2015, MNRAS, 448, 684.


\bibitem[Smallwood et al.(2018)]{smaetal2018} Smallwood, J.~L., Martin, R.~G., Livio, M., \& Lubow, S.~H.\ 2018, MNRAS, 480, 57

\bibitem[Soker(1998)]{soker1998} Soker, N.\ 1998, AJ, 116, 1308 

\bibitem[Staff et al.(2016)]{staetal2016} Staff, J.~E., De Marco, O., Wood, P., Galaviz, P., \& Passy, J.-C.\ 2016, MNRAS, 458, 832 


\bibitem[Stone et al.(2015)]{stoetal2015} Stone, N., Metzger, B.~D., \& Loeb, A.\ 2015, MNRAS, 448, 188 

\bibitem[Tadeu dos Santos et al.(2015)]{dosetal2015} Tadeu dos Santos, M., Correa-Otto, J.~A., Michtchenko, T.~A., Ferraz-Mello, S.\ 2015, A\&A, 573, A94.

\bibitem[Terquem \& Papaloizou(2007)]{terpap2007} Terquem, C. \& Papaloizou, J.~C.~B.\ 2007, ApJ, 654, 1110.

\bibitem[Teyssandier \& Terquem(2014)]{teyter2014} Teyssandier, J., \& Terquem, C.\ 2014, MNRAS, 443, 568

\bibitem[Tremblay et al.(2016)]{treetal2016} Tremblay, P.-E., Cummings, J., Kalirai, J.~S., et al.\ 2016, MNRAS, 461, 2100 
  


\bibitem[Vanderburg et al.(2015)]{vanetal2015} Vanderburg, A., Johnson, J.~A., Rappaport, S., et al.\ 2015, Nature, 526, 546 


\bibitem[Veras et al.(2011)]{veretal2011} Veras, D., Wyatt, M.~C., Mustill, A.~J., Bonsor, A., \& Eldridge, J.~J.\ 2011, MNRAS, 417, 2104 



\bibitem[Veras et al.(2013a)]{veretal2013a} Veras, D., Hadjidemetriou, J.~D., \& Tout, C.~A.\ 2013a, MNRAS, 435, 2416 

\bibitem[Veras et al.(2013b)]{veretal2013b} Veras, D., Mustill, A.~J., Bonsor, A., \& Wyatt, M.~C.\ 2013b, MNRAS, 431, 1686 

\bibitem[Veras \& Evans(2013)]{vereva2013} Veras, D., \& Evans, N.~W.\ 2013, MNRAS, 430, 403 

 


\bibitem[Veras et al.(2014a)]{veretal2014a} Veras, D., Jacobson, S.~A., G\"{a}nsicke, B.~T.\ 2014a, MNRAS, 445, 2794 

\bibitem[Veras et al.(2014b)]{veretal2014b} Veras, D., Leinhardt, Z.~M., Bonsor, A., G\"{a}nsicke, B.~T.\ 2014b, MNRAS, 445, 2244

\bibitem[Veras et al.(2014c)]{veretal2014c} Veras, D., Evans, N.~W., Wyatt, M.~C., \& Tout, C.~A.\ 2014c, MNRAS, 437, 1127 

\bibitem[Veras et al.(2014d)]{veretal2014d} Veras, D., Shannon, A., \& G{\"a}nsicke, B.~T.\ 2014d, MNRAS, 445, 4175 

\bibitem[Veras \& G\"{a}nsicke(2015)]{vergae2015} Veras, D., G\"{a}nsicke, B.~T.\ 2015, MNRAS, 447, 1049 


\bibitem[Veras et al.(2015)]{veretal2015} Veras, D., Eggl, S., G{\"a}nsicke, B.~T.\ 2015, MNRAS, 451, 2814 


\bibitem[Veras(2016a)]{veras2016a} Veras, D.\ 2016a, Royal Society Open Science, 3, 150571 

\bibitem[Veras(2016b)]{veras2016b} Veras, D.\ 2016b, MNRAS, 463, 2958 

\bibitem[Veras et al.(2016)]{veretal2016} Veras, D., Mustill, A.~J., G{\"a}nsicke, B.~T., et al.\ 2016, MNRAS, 458, 3942 



\bibitem[Veras et al.(2018)]{veretal2018} Veras, D., Georgakarakos, N., G{\"a}nsicke, B.~T., \& Dobbs-Dixon, I.\ 2018, MNRAS, 481, 2180


\bibitem[Villaver \& Livio(2007)]{villiv2007} Villaver, E., \& Livio, M.\ 2007, ApJ, 661, 1192 

\bibitem[Villaver \& Livio(2009)]{villiv2009} Villaver, E., \& Livio, M.\ 2009, ApJL, 705, L81 

\bibitem[Villaver et al.(2014)]{viletal2014} Villaver, E., Livio, M., Mustill, A.~J., \& Siess, L.\ 2014, ApJ, 794, 3

\bibitem[Vokrouhlicky(1998)]{vokrouhlicky1998} Vokrouhlicky, D.\ 1998, A\&A, 335, 1093  

\bibitem[Vokrouhlick{\'y} et al.(2015)]{voketal2015} Vokrouhlick{\'y}, D., Bottke, W.~F., Chesley, S.~R., Scheeres, D.~J., \& Statler, T.~S.\ 2015, In Asteroids IV (Eds: Patrick Michel, Francesca E. DeMeo, and William F. Bottke), Pgs. 509-531 

\bibitem[Vokrouhlick{\'y} et al.(2007)]{voketal2007} Vokrouhlick{\'y}, D., Breiter, S., Nesvorn{\'y}, D., \& Bottke, W.~F.\ 2007, Icarus, 191, 636 

\bibitem[Voyatzis et al.(2013)]{voyetal2013} Voyatzis, G., Hadjidemetriou, J.~D., Veras, D., \& Varvoglis, H.\ 2013, MNRAS, 430, 3383 

\bibitem[Wang \& Hou(2017)]{wanhou2017} Wang, X., \& Hou, X.\ 2017, MNRAS, 471, 243

\bibitem[Wickramasinghe et al.(2010)]{wicetal2010} Wickramasinghe, D.~T., Farihi, J., Tout, C.~A., Ferrario, L., \& Stancliffe, R.~J.\ 2010, MNRAS, 404, 1984 



\bibitem[Wilson et al.(2015)]{wiletal2015} Wilson, D.~J., G{\"a}nsicke, B.~T., Koester, D., et al.\ 2015, MNRAS, 451, 3237 

\bibitem[Wilson et al.(2016)]{wiletal2016} Wilson, D.~J., G{\"a}nsicke, B.~T., Farihi, J., \& Koester, D.\ 2016, MNRAS, 459, 3282 

\bibitem[Wisdom(1980)]{wisdom1980} Wisdom, J.\ 1980, AJ, 85, 1122.
  



\bibitem[Xu et al.(2013)]{xuetal2013} Xu, S., Jura, M., Klein, B., Koester, D., \& Zuckerman, B.\ 2013, ApJ, 766, 132 

\bibitem[Xu et al.(2014)]{xuetal2014} Xu, S., Jura, M., Koester, D., Klein, B., \& Zuckerman, B.\ 2014, ApJ, 783, 79 



\bibitem[Xu et al.(2017)]{xuetal2017} Xu, S., Zuckerman, B., Dufour, P., et al.\ 2017, ApJL, 836, L7 

\bibitem[Xu et al.(2018)]{xuetal2018} Xu, W., Lai, D., \& Morbidelli, A.\ 2018, arXiv:1805.07501  




\bibitem[Zuckerman et al.(2003)]{zucetal2003} Zuckerman, B., Koester, D., Reid, I.~N., H\"{u}nsch, M.\ 2003, ApJ, 596, 477 

\bibitem[Zuckerman et al.(2010)]{zucetal2010} Zuckerman, B., Melis, C., Klein, B., Koester, D., \& Jura, M.\ 2010, ApJ, 722, 725 



\end{thebibliography}
\end{document}